\documentclass[aps,pre,showpacs,preprintnumbers,twocolumn,superscriptaddress]{revtex4-2}

\usepackage{graphicx}
\usepackage{dcolumn}
\usepackage{bm}
\usepackage{hyperref}
\hypersetup{
    colorlinks=true,
    linkcolor=blue,
    citecolor=blue,
    urlcolor=blue
}
\usepackage{mathtools}
\newcommand{\norm}[1]{\left\lVert#1\right\rVert}
\newcommand{\Matrix}[1]{\mathbf{#1}}
\usepackage{float}
\usepackage{ulem}
\usepackage{comment}

\newcommand{\lcc}[1]{\textcolor{blue}{[#1]}}

\begin{document}

\title{Odd pathways speed up self-assembly}

\author{Dawid Dopierała}
\affiliation{TCM group, Cavendish Laboratory, University of Cambridge, Cambridge, UK}

\author{Luca Cocconi}
\affiliation{TCM group, Cavendish Laboratory, University of Cambridge, Cambridge, UK}

\author{Robert L.~Jack}
\affiliation{DAMTP, University of Cambridge, Cambridge, UK}
\affiliation{Yusuf Hamied Department of Chemistry, University of Cambridge, UK}

\author{Anton Souslov}
\affiliation{TCM group, Cavendish Laboratory, University of Cambridge, Cambridge, UK}

\date{\today}

\begin{abstract}
Active self-assembly can bypass equilibrium bottlenecks through external energy injection. However, generic driving typically distorts target structures and requires sustained energy input even after assembly is complete. Here, we investigate a class of non-reciprocal interactions that accelerates assembly while preserving the equilibrium Boltzmann distribution. The probability currents induced by these odd interactions reshape fundamental processes, including activated barrier crossing, soft-mode relaxation, and transitions between metastable states. In particular, these currents enhance Arrhenius rates by driving particles across otherwise inaccessible free-energy barriers. We show that this acceleration arises from an effective increase in the mobility of the reaction coordinate, mediated by non-reciprocal coupling between mechanical modes. In turn, we discover a trade-off between kinetic acceleration and power dissipation when active forces are engaged. Our results suggest a route to energy-efficient, high-fidelity self-assembly via active catalysts that transiently accelerate relaxation toward equilibrium targets and deactivate upon reaching the desired state.
\end{abstract}

\maketitle

Self-assembly is a scalable and energy efficient approach for forming complex structures at the nanoscale~\cite{grzybowski2009self, hagan2021equilibrium}. However, many metastable intermediates and kinetic traps in the configurational energy landscape obstruct the realisation of target structures. At thermodynamic equilibrium, these barriers can only be overcome through rare thermally activated fluctuations~\cite{hanggi1990reaction, whitelam2015statistical}. By contrast, active self-assembly promises a cure for this {\it Kramers bottleneck} using sustained probability currents~\cite{mallory2018active}. This idea has led to a number of promising demonstrations of activity-assisted assembly: for example, self-propulsion has been employed to accelerate the formation of two-dimensional Kagome lattices~\cite{schubert2025self} and to enhance the annealing of colloidal monolayers using active intruder particles~\cite{ramananarivo2019activity}. 
 
In these and related examples, activity perturbs the steady state, and there is no guarantee that the target configuration will be preserved~\cite{martin2021statistical, fruchart2021non}.  Even if active self-assembly leads to the target configuration, the system might require continuous energy input to avoid disassembly~\cite{soto2014self}. An approach where  activity  instead leads to equilibrium structures would open the way to fast, robust self-assembly without the need for sustained energy injection once assembly completes.

Like other types of activity, {\it non-reciprocal} interactions---a specific class of nonequilibrium couplings that violate action-reaction symmetry---can speed up diffusive dynamics by driving the system away from equilibrium~\cite{fruchart2021non}. Moreover, it is possible to add non-reciprocal forces without distorting the equilibrium probability distribution~\cite{hwang2005accelerating, kaiser2017acceleration}, as recently demonstrated for glassy systems \cite{ghimenti2023sampling, ghimenti2024transverse}. However, in the context of self-assembly, non-reciprocal interactions have primarily been exploited to engineer novel steady states, such as cyclic far-from-equilibrium dynamics~\cite{osat2023non, osat2024escaping, metson2025continuous}. Here, we ask a complementary question: how does one harness non-equilibrium dynamics to accelerate self-assembly to an {\it equilibrium} conformation?

We show that a broad class of non-reciprocal forces (referred to as transverse forces)
speeds up assembly dynamics while leaving the equilibrium distribution unchanged.
We derive analytical and numerical results for active speed up and its energetic costs in minimal models of barrier crossings and folding of complex shapes. 
These findings establish a simple strategy for controlled assembly: initially, non-reciprocal forces are activated to drive otherwise rare transitions, and then deactivated once the target state is reached. Such an approach applies to systems as diverse as materials assembly, cargo binding, and synthetic catalysts. 
More generally, we introduce a mechanism  by which transverse forces control currents independently of probability distributions in active self-assembly.

\begin{figure*}[!ht]
    \centering
    \includegraphics[]{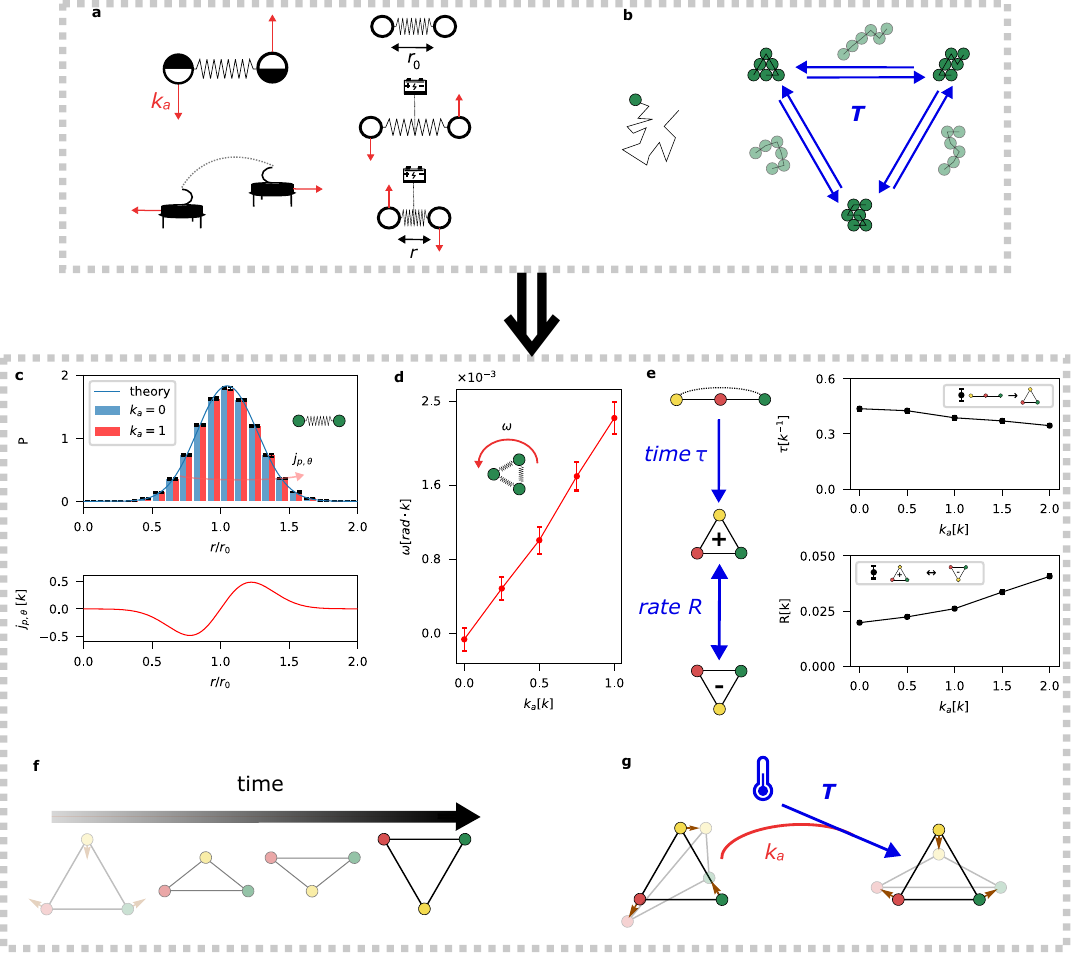}
    \caption{\textbf{Transverse forces as effective temperature.} How do transverse forces affect self-assembly? \textbf{a}, Abstraction of transverse forces: pairs of metallodielectric (Janus) colloids driven by a uniform AC electric field~\cite{boymelgreen2014spinning}, Kilobots~\cite{rubenstein2014kilobot}, and ball-and-spring model. Battery represents energy injection, which results in transverse forces (red arrows) proportional to $k_a(r-r_0)$. \textbf{b}, Abstraction of self-assembly:
    thermal noise leads to the assembly of target structures, for example in colloidomers~\cite{mcmullen2022self}.  Resulting phenomenology: \textbf{c},  Steady-state probability distribution for two-particle system (top, inset) remains the same with and without transverse forces (top). However, transverse forces introduce probability currents $j_{p,\theta}$ (bottom). ($T=0.05$) \textbf{d},  Adding transverse forces to a triangle (inset) causes a net rotation with mean angular velocity $\omega$ set by activity $k_a$. ($T=0.001$) \textbf{e}, By adding transverse forces, the assembly time $\tau$ of a triangle from an initial line configuration decreases (top), and transition rate $R$ between triangles with different chiralities (enantiomers) increases. ($T=0.05$) The mechanism of this activity-induced acceleration involves coupling the eigenmode that facilitates each flip, \textbf{f}, to the fluctuations in a second mode via transverse forces, \textbf{g}.
    }
    \label{Fig1}
\end{figure*}
We consider self-assembly in the specific context of mechanical networks of overdamped thermal colloids in the presence of nonconservative forces. The colloid dynamics are governed by the Langevin equation:
\begin{equation}
\dot{\bm{r}}_i=- \left[ \bm\nabla_i  + c_i\bm\nabla^\perp_i \right] E
  + \sqrt{2T} \bm{\eta}_i,
\label{model}
\end{equation} where we refer to $E(\{\bm{r}_i\})$ as the energy of the system, \mbox{$\bm\nabla^\perp \equiv \epsilon_{kl} \partial_l$} is the two-dimensional curl with $\epsilon_{kl}$ the Levi-Civita symbol. Physically, the transverse force $\bm\nabla^\perp E$ (illustrated in Fig.~\ref{Fig1}\textbf{a} and also known as an odd~\cite{scheibner2020odd, yasuda2021odd, fruchart2023odd} or curl~\cite{wu2009direct, sun2009brownian,berry2015hamiltonian}  force) breaks reciprocity of mechanical action and reaction. These transverse forces can arise from, for example, hydrodynamic interactions~\cite{tan2022odd, bililign2022motile} or be programmed using robotic components~\cite{brandenbourger2019non,binysh2026more}. We take the prefactor $c_i = c$ to be a constant, where for harmonic $E(\bm{r})$, we define the active spring constant $k_a \equiv c k$ (see Methods~\ref{Langevin}). In many-body systems, such forces give rise to out-of-equilibrium emergent properties, including odd elasticity~\cite{scheibner2020odd, yasuda2021odd, tan2022odd, bililign2022motile,fruchart2023odd}. Alongside these transverse forces, Eq.~(\ref{model}) introduces thermal fluctuations, represented by the Gaussian white noise $\eta$, which enable spontaneous transitions between metastable configurations and facilitate self-assembly along accessible pathways, see Fig.~\ref{Fig1}\textbf{b}.

A remarkable feature of these dynamics is that the stationary probability distribution of the particle coordinates is Boltzmann-like, i.e., $p \sim e^{-E(\{\bm{r}_i\})/k_B T}$, and independent of the odd force parameter $c_i$ (see Methods~\ref{Langevin} for a proof).
The origin of this invariance is geometric: the odd force is everywhere orthogonal to the energy gradient and generates divergence-free probability currents that circulate along constant-energy manifolds without redistributing probability between them~\cite{ghimenti2023sampling, o2024geometric, ghimenti2024transverse}. Nevertheless, odd forces reshape dynamical pathways, leading to the appearance of nonvanishing probability currents in configuration space when $k_a \neq 0$. 

The simplest model to illustrate the appearance of currents within a Boltzmann distribution is the two-particle odd spring, defined in Methods~\ref{sec:models-methods}. In this minimal setting, circulating probability currents along constant-energy manifolds appear whenever the spring is away from its rest length, and do not alter the stationary distribution of configurations (Fig.~\ref{Fig1}\textbf{c}). Further details are provided in Methods~\ref{Two-particle model (arbitrary $T$)}.

\subsection*{Odd triangle: signatures of non-equilibrium dynamics}
Another minimal example is a three-particle network connected via odd springs, as defined in Methods~\ref{sec:models-methods}. In this geometry, the introduction of transverse forces induces slow global rotations (consistent with the previous studies~\cite{tan2022odd}), quantified in Fig.~\ref{Fig1}\textbf{d}. Unlike athermal odd-elastic materials, here global rotations arise in the absence of an externally applied stress. Instead, thermal fluctuations transiently stretch the triangle, and the curl forces convert these entropically favoured configurations into a non-zero average torque on each spring. Active forces also manifest in the acceleration of activated barrier-crossing processes, corresponding to mirror-symmetry flips between enantiomers characterised by the different sign of their signed area $A$ (Fig.~\ref{Fig1}\textbf{e}, flip rate $R$ estimation procedure described in Methods~\ref{Flip rates estimation}). For the non-reciprocal coefficient $k_a$ of order unity, the flip rate $R$ increases by approximately a factor of two.

The enhancement of the flip rate $R$ can be traced to a mode-coupling mechanism induced by odd forces. To make this mechanism explicit, we consider the low-temperature eigenmodes of the purely reciprocal ($k_a=0$) system. In this limit, the eigenmodes are dynamically decoupled. Among these, a particular deformation plays a central role in facilitating activated transitions, as shown in Fig.~\ref{Fig1}\textbf{f}. This mode corresponds to a slight separation of two particles, which transiently opens space for the third particle to pass between them. Transverse interactions couple the equilibrium eigenmodes, effectively leading to mode mixing and to the injection of additional noise contributions into the kinetics responsible for flipping. This mechanism is illustrated in Fig.~\ref{Fig1}\textbf{e} and discussed in Supplementary Information.

More generally, the triangle model provides a minimal coarse-grained representation of non-equilibrium structural rearrangements, isolating how non-reciprocal forces reshape dynamical pathways. In this sense, triangle flips exemplify a mechanism analogous to the reconfiguration of nucleation kernels in driven and active materials~\cite{tjhung2018cluster}. This suggests a broader principle: transverse forces modify collective dynamics by coupling equilibrium eigenmodes and effectively renormalising the noise acting along dominant transition pathways. To test how this mechanism operates beyond fully rigid structures, we now study a system in which rigid and soft collective modes coexist.
\subsection*{Odd folding through a soft mode}
\begin{figure*}[!ht]
    \centering
    \includegraphics[]{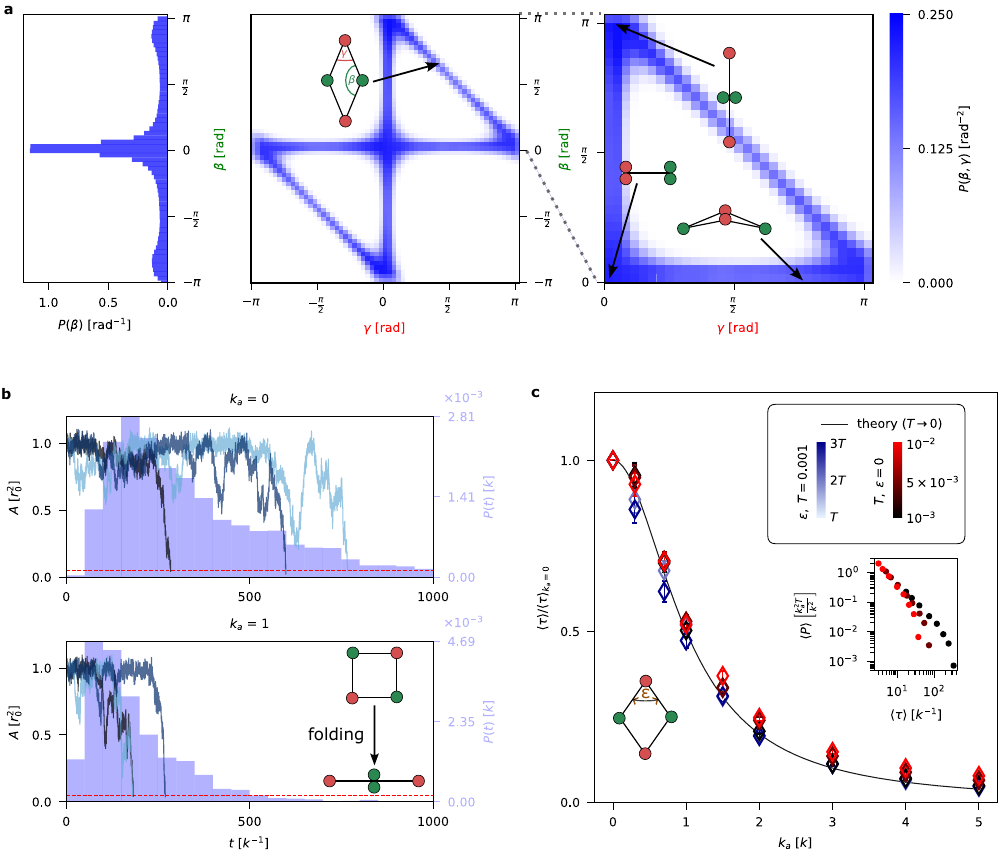} 
    \caption{\textbf{Activity-enhanced folding.} \textbf{a},  Steady-state probability distribution of the angles $\beta$ and $\gamma$ for a rhombic configuration of four springs. Horizontal and vertical lines indicate transitions via wedge states, while diagonals correspond to rhombus configurations. Zoom of the first quadrant (right) shows that, starting from the square configuration (centre of the diagonal), the system first assembles into a line with the length of two springs. In the steady state, the system is biased towards near-zero $\beta$ configurations (left). ($T=0.001$, $k_a=0$) \textbf{b}, Sample trajectories of the system signed area $A$ and histograms of folding times over many runs (shaded). Transverse forces reduce the folding time; the red line indicates the folding threshold. ($T=0.001$)  \textbf{c}, Simulations confirm the scaling of the folding time $\tau$ with the activity coefficient $k_a$ predicted by Eq.~\eqref{beta dynamics}, both without and with ($\varepsilon \neq 0$) angular potential. The inset shows the power as a function of the folding time, highlighting that faster folding is associated with higher energy injection.}
    \label{Fig2}
\end{figure*}

Consider a floppy square consisting of four particles connected by four springs, as defined in Methods~\ref{sec:models-methods}. This motif folds through a sequence of rhombus-like configurations connecting an extended line state to a compact square, as illustrated in Fig.~\ref{Fig2}\textbf{a}. While all rhombus configurations are energetically equivalent, the line state is entropically stabilized through an order-by-disorder mechanism, since fluctuations around the line access a larger volume of microstates than those around the rhombus states~\cite{rocklin2018folding}. 

To analyse this system, we project the steady-state probability distribution onto the two-dimensional configuration subspace defined by the internal angles $\beta$ and $\gamma$ (Fig.~\ref{Fig2}\textbf{a}). This data confirms that the zero-energy rhombus-like pathway forms the dominant folding channel connecting line and square states.
However, we observe that upon introducing transverse forces, the dynamics along the soft mode are accelerated. 
Sample trajectories of the signed area $A$ of the system (Fig.~\ref{Fig2}\textbf{b}) show that the system traverses the rhombus pathway more rapidly, reaching the line configuration faster than in the purely reciprocal case.

To study the folding process quantitatively, we derive a reduced description of the stochastic dynamics of the internal angle $\beta \in [0,\pi]$ parametrising the floppy manifold of the square network, where $\gamma=\pi-\beta$ (Fig.~\ref{Fig2}a). Here, $\beta = \pi/2$ and $\beta = \{0,\pi\}$ correspond to the compact square and the two-line configurations, respectively. Since all other modes are massive and relax over timescales of order of the inverse elastic coefficient, these modes are treated as fast degrees of freedom and thus formally integrated out to obtain the effective Langevin equation:
\begin{equation}
    \dot{\beta} = -2T \Lambda\cot{\beta}+ 2\sqrt{T \Lambda} \eta,
    \label{beta dynamics}
\end{equation}
where:
\begin{equation}
    \Lambda =1+\left (\frac{k_a}{k}\right)^2
    \label{eq:Lambda}
\end{equation}
is a dimensionless speed-up factor stemming from the presence of non-reciprocal forces. Importantly, $\Lambda$ multiplies both the deterministic drift and the noise amplitude, leaving the steady-state distribution unchanged while uniformly accelerating the kinetics. The drift term has a purely entropic origin, as evidenced by its vanishing in the zero-temperature limit, and diverges as $\beta \to \{0, \pi\}$. This divergence is a consequence of the harmonic approximation to the spring potential used in the low-temperature expansion~\cite{rocklin2018folding}. In the trajectories shown in Fig.~\ref{Fig2}\textbf{b}, this appears as a sudden drop in the system's area $A$ preceding full folding. Significantly, the speed-up factor $\Lambda$ scales as 
$\left (k_a/k\right)^2$ for large $k_a/k$, suggesting that in principle an arbitrary speed up could be possible.

The detailed derivation of Eq.~\eqref{beta dynamics} (see Methods \ref{Derivation of the reduced beta dynamics} and Supplementary Information) involves writing the full Fokker-Planck equation for the particle coordinates in curvilinear variables that separate two rigid translations ($x_\textrm{cm}, y_\textrm{cm}$), a global rotation ($\theta$), the soft bond-bending mode ($\beta$), and four orthogonal shear deformations ($\alpha_{k \in \{c,d,e,f\}}$). Expanding the drift terms up to linear order in small deformations and for fixed $\beta$, one finds that all fast modes acquire a harmonic restoring force of order $k$ and relax rapidly to a Gaussian distribution around their conditional mean. We arrive at this fast-slow limit by showing that the stiff mode $\alpha_c$ operates as Gaussian white noise in the $\beta$ dynamics.  Hence, the activity renormalizes both the deterministic drift and the additive noise strength by the same factor $\Lambda$.

To test the robustness of this speed-up mechanism, we included a weak interaction acting at each internal angle $\psi_i$, described by the potential:
\begin{equation}
    V_{\text{angle}}({\psi_i})= -\frac{\varepsilon}{4}\cos^2\!\left(2{\psi_i}\right),
    \label{eq:V-angle}
\end{equation}
which for $\varepsilon\sim T$ weakly stiffens the folding mode. These angle interactions act as small impurities along the otherwise flat folding pathway, selecting the preferred folded configurations. After a similar coarse-graining procedure (see Methods~\ref{Derivation of the reduced beta dynamics} and Supplementary Information), we obtain a variant of Eq.~\eqref{beta dynamics} (Eq.~\eqref{eq:beta_sde_methods} in Methods) augmented with a drift term $-4 \varepsilon \Lambda \sin (4 \beta)$, consistent with the measure-preserving acceleration discussed above. 

The reduced dynamics in Eq.~\eqref{beta dynamics} make a simple quantitative prediction for how the non-reciprocal force accelerates folding along the soft rhombus mode, namely that transition rates with and without non-reciprocal forces should be related by the multiplicative factor $\Lambda$. We test this prediction in Fig.~\ref{Fig2}\textbf{c} against numerical simulations of the full system with and without bond-bending forces. We observe a remarkable agreement across a broad range of temperatures $T$, interaction strengths $\varepsilon$, and transverse couplings $k_a$, despite the assumptions underlying the derivation, including time-scale separation, a low-temperature expansion, and the implicit assumption that folding proceeds through the same soft pathway for both $\varepsilon=0$ and $\varepsilon\neq0$. 

Our numerical simulations also allow us to quantify the thermodynamic cost of nonequilibrium acceleration. In particular, Fig.~\ref{Fig2}\textbf{c} (inset) shows how the mean power dissipated by the transverse forces during folding ($P$) varies with the mean assembly time ($\tau$). The data reveal a clear trade-off: although self-assembly can be sped arbitrarily, this gain is systematically accompanied by higher dissipation, whereas constraining the thermodynamic budget directly limits the achievable acceleration.
\subsection*{Cargo-assisted stabilization of metastable states}
The square folding provided an analytically tractable model of how non-reciprocal forces accelerate folding along a dominant pathway. Its steady-state structure is simple: the square evolves only through a one-parameter sequence of zero (or near-zero) energy configurations, and activity only controls the time required to reach the most entropically favoured states. In many realistic self-assembling systems, by contrast, multiple long-lived configurations and reconfiguration pathways coexist, and functionality depends not only on assembly speed but on controlling transitions between competing states~\cite{whitelam2015statistical}.

\begin{figure*}[!ht]
    \centering
    \includegraphics[]{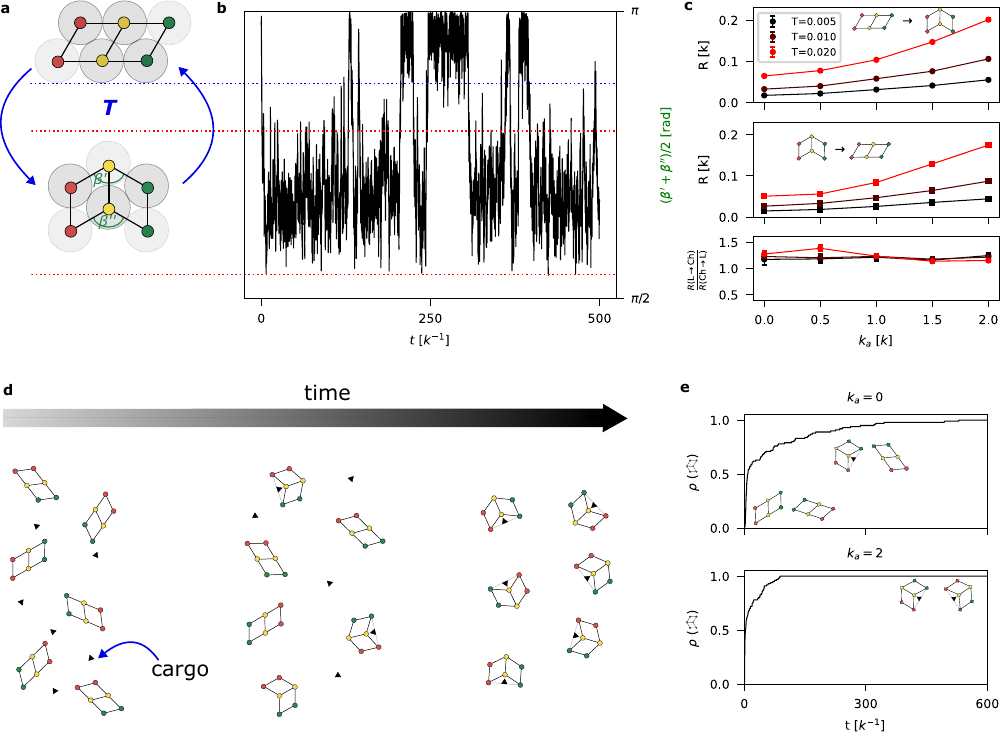}
    \caption{\textbf{Acceleration of transitions between metastable states.} \textbf{a}, Ladder (top) and chevron (bottom) are metastable configurations of the double-square system enforced by the Morse interaction (halos indicate interaction range; faded halos denote effectively non-interacting particles). A sample trajectory (\textbf{b}, $k_a=2$, $T=0.005$) shows repeated transitions between ladder (above the blue dotted line) and chevron (between the red dotted lines). \textbf{c}, Transverse forces increase the transition rates $R$ for ladder$\to$chevron (top) and chevron$\to$ladder (middle), while their ratio remains constant (bottom), indicating probability invariance. \textbf{d}, Accelerated stabilization in a system of non-interacting double-squares. Cargo particles temporarily bind to the chevron configuration, kinetically trapping it. \textbf{e}, ($T=0.005$) Sample trajectory of the chevron density $\rho$ in the non-interacting system with cargo. Activity enhances independent flips, leading to faster stabilization in the chevron-only state (bottom) compared to the passive case (top). 
    }
    \label{Fig3}
\end{figure*}

To capture this more complex regime, we now turn to a six-particle network, the double square shown in Fig.~\ref{Fig3}\textbf{a}, which represents the minimal extension of the square that exhibits genuine metastability. In this system, two distinct shapes, a chevron and a ladder, coexist, with different equilibrium probabilities determined by the free energy landscape. The double square thus provides a natural toy model to demonstrate how odd forces can enhance transition rates between genuinely metastable states without biasing their steady-state populations.

We model this system using Eq.~\eqref{model} with harmonic interactions between nearest neighbours and an additional short-range Morse attraction between all pairs of particles that are not directly connected by a spring. The equilibrium distance of the Morse interaction is matched to the spring rest length in order to stabilise the target configurations against thermal noise (see Methods~\ref{sec:models-methods} for further details). Figure~\ref{Fig3}\textbf{b} shows a representative trajectory of the average internal angles for a single network, highlighting intermittency between the metastable states. Thresholds for classifying the instantaneous configuration (dashed lines) provide an operational definition for counting transition events. Averaging over long trajectories, we once again find a speed-up in the corresponding transition rates with the odd coefficient $k_a/k$ (Fig.~\ref{Fig3}\textbf{c}, flip rate $R$ estimation procedure described in Methods~\ref{Flip rates estimation}), while their ratio, which controls the relative occupation of the two states, remains unaffected (Fig.~\ref{Fig3}\textbf{d}).

The independent enhancement of transition rates suggests a novel mechanism for controlling assembly at the system level. Figure~\ref{Fig3}\textbf{d} illustrates a kinetic stabilization strategy in which transient cargo particles selectively bind to chevron configurations, effectively \textit{locking in} the desired state. Because transverse forces accelerate transitions without biasing the steady-state probabilities, the system efficiently explores configuration space, rapidly populating chevron states until cargo binding arrests them.
Simulations of a macroscopic assembly of non-interacting double-square particles confirm the impact of this mechanism. As shown in Fig.~\ref{Fig3}\textbf{e}, activity dramatically shortens the time required to stabilize the entire population in the target chevron configuration, achieving complete assembly faster than in the absence of activity (see  Supplementary Information for simulation details). Transverse forces therefore act as a system-wide kinetic amplifier, enabling fast, reliable, and equilibrium-preserving control of collective self-assembly.

\subsection*{Towards faster bioinspired self-assembly}
We have presented a general mechanism for controlling assembly kinetics independently of equilibrium properties by introducing non-reciprocal forces. Our minimal models quantify the achievable speedup and energetic cost of key assembly pathways, from shape transitions to folding. More broadly, our results demonstrate that active self-assembly exploits a general principle: transverse forces can generate probability currents without altering the underlying equilibrium distribution.
This decoupling of kinetics from thermodynamics may provide a useful strategy for accelerating self-assembly while preserving target structures. Because self-assembly underlies systems ranging from molecular complexes such as proteins~\cite{onuchic1997theory, dill2012protein} and viral capsids~\cite{hagan2006dynamic, perlmutter2015mechanisms} to engineered materials based on colloids~\cite{glotzer2007anisotropy, walther2008janus, sacanna2010lock}, colloidomers~\cite{mcmullen2022self}, DNA origami~\cite{rothemund2006folding, ji2021dna} and nanorobotics~\cite{palagi2018bioinspired, law2023micro}, this control of assembly pathways could enable the design of robust self-organizing systems operating under realistic energetic constraints.

\begin{acknowledgments}
D.D.~acknowledges funding from the Cambridge Trust \& Robinson College Corbridge Studentship. L.C.~acknowledges funding from the Ernest Oppenheimer Fund.
A.S.~acknowledges funding from UKRI under project  UKRI3173. This research was supported in part by grant NSF PHY-2309135 to the Kavli Institute for Theoretical Physics (KITP).
\end{acknowledgments}

\bibliography{paper_bib_fixed}

\clearpage
\appendix
\renewcommand\thefigure{A\arabic{figure}}
\setcounter{figure}{0}
\parskip1pt

\section{Methods}
\subsection{Langevin dynamics and steady-state probability distribution invariance}
\label{Langevin}

\definecolor{mygreen}{rgb}{0.0,0.55,0.3}
\newcommand{\rlj}[1]{{\color{mygreen}#1}}

We consider a minimal stochastic model that isolates non-reciprocal mechanical response. Each particle experiences a force composed of three contributions: (i) a reciprocal part derived from a potential $E$, (ii) a transverse, non-reciprocal odd-elastic component, and (iii) additive isotropic Gaussian noise representing thermal fluctuations. We work in the overdamped regime, appropriate for colloidal, soft-matter, and robotic systems where inertia is negligible.

The position of particle $i$ is $\bm{r}_i=(x_i,y_i)$.  For systems of odd springs, the interaction energy is given by harmonic potentials with spring constant $k$ and natural length $r_0$, as
\begin{equation} 
    E = \sum_{\langle ij\rangle } \frac{k(\norm{\bm{r}_j-\bm{r}_i}-r_0)^2}{2},
\end{equation}
where the sum runs over distinct pairs of connected particles. 
The strength of the transverse coupling is $k_a$, and the resulting dynamics is
\begin{equation}
    \dot{\bm{r}}_i= \bm{F}_i 
 + \sqrt{2T} \bm{\eta}_i,
    \label{model_methods}
\end{equation}
with the force on particle $i$ is
\begin{equation}
\bm{F}_i  =   -\left[ \Matrix{I}+\frac{k_a}{k} 
    \begin{pmatrix}
0 & 1 \\
-1 & 0
\end{pmatrix} \right ]  \nabla_i E
\label{eq:Fi}
\end{equation}
similar to Eq.~\eqref{model} of the main text; here $T$ is the noise strength (which we can identify with the temperature, in appropriate units); also $\langle \eta_{i \alpha} (t) \eta_{j \beta} (t') \rangle = \delta_{ij} \delta_{\alpha \beta} \delta(t-t')$. Numerical solution of these equations is discussed in Section~\ref{sec:methods-numerics}, along with other numerical methods.

Note that the transverse force (proportional to $k_a$)is everywhere orthogonal to the gradient of $E$, providing an antisymmetric, non-conservative contribution to the dynamics, whose consequences we discuss below.  

\paragraph{Invariance of steady-state probability distribution.}
For $k_a=0$ the steady-state probability distribution for Eq.~\eqref{model_methods} is the Boltzmann distribution 
$p^\star \propto e^{-E/T}$ and this steady state has time-reversal symmetry (in particular there are no steady-state currents).  We now show that introducing $k_a\neq 0$ does not affect the steady state distribution, although it does lead to finite currents in the steady state, so time-reversal symmetry is broken.  Moreover, the system retains a non-trivial time-reversal symmetry associated with reversing the sign of $k_a$.

We derive this following the condition from~\cite{hwang2005accelerating}.
Consider the Ito stochastic differential equation of a 2D system with $M$ particles, whose state is 
$\bm{X}=(x_1, y_1, x_2, y_2,\dots,x_M,y_M)$:
\begin{equation}
\dot{\bm{X}} = -\nabla E(\bm{X}) + \mathbf Y(\bm{X}) + \sqrt{2T}\,\boldsymbol\eta(t),
\end{equation}
where $(x_i,y_i)$ is the position of the $i$-th particle, as above; also $E$ is the energy and the vector field $\mathbf{Y}$ describes a non-reciprocal interaction is not the gradient of any scalar function. The corresponding Fokker-Planck equation for the probability density $p(\bm{x},t)$ is:
\begin{equation}
\partial_t p = \nabla \cdot \left[(\nabla E - \mathbf Y)p + T\nabla p\right].
\label{fokker planck}
\end{equation} 
A necessary and sufficient condition for the Boltzmann density $p^\star \sim e^{-E/T}$ to be unchanged by the addition of the non-reciprocal interaction $\bm{Y}$ is $\nabla \cdot ( \bm{Y} e^{-E/T})=0$,
which is equivalent to:
\begin{equation}
\nabla\cdot\bm{Y} = \frac{1}{T}\bm{Y}\cdot\nabla E.
\label{steady_condition}
\end{equation}

For the class of odd-elastic interactions studied in the paper, the non-conservative contribution can be written as:
\begin{equation}
\bm{Y}(\bm{x}) = -\frac{k_a}{k}\Matrix{\mathcal E} \nabla E(\bm{X}),
\label{eq:Ycond}
\end{equation}
where the matrix $\Matrix{\mathcal E}$ has a block diagonal form consisting of $M$ copies of the  $2\times2$ Levi-Civita matrix [which is the anti-symmetric $2\times2$ matrix appearing in Eq.~\eqref{model_methods}]. Putting this form for $\bm{Y}$ into the two sides of~Eq.~\eqref{steady_condition} gives:
\begin{align}
\nabla\cdot\mathbf Y &= -\frac{k_a}{k}\sum_{i,j}\mathcal E_{ij}\partial_i\partial_j E,\\
\mathbf Y\cdot\nabla E &= -\frac{k_a}{k}(\nabla E)^{T}\boldsymbol{\mathcal E}\nabla E.
\end{align}
Both of these terms vanish due to the anti-symmetry of $\Matrix{\mathcal E}$, so Eq.~\eqref{eq:Ycond} holds, and the Boltzmann distribution $p^\star$ is an exact stationary solution of~\eqref{fokker planck}, independent of $k_a$. Note this result uses only that $\Matrix{\mathcal E}$ is anti-symmetric and independent of $\bm{x}$.

\paragraph{Probability currents.}
Although the stationary density is independent of $k_a$, the odd-elastic forces generically produce nonzero probability currents. Substituting $p^\star$ into the probability flux $\bm{j}_\mathrm{p}=(\bm{Y}-\nabla E)p - T\nabla p$ yields:
\begin{equation}
\bm{j}_\mathrm{p}^\star(\bm{X}) = \bm{Y}(\bm{X})\,p^\star(\bm{X}) = -\frac{k_a}{k}p^\star(\bm{X}) \boldsymbol{\mathcal E}\nabla E(\bm{X}),
\end{equation}
This current is divergence-free, as it must be because $p^\star$ is the steady state.  The fact that $\bm{j}_\mathrm{p}^\star$ is finite signals the breaking of time-reversal symmetry, see also below~\cite{seifert2012stochastic}.  Note however that reversing the sign of $k_a$ corresponds to multiplying $\bm{j}_\mathrm{p}^\star$ by $-1$.  This means that the time-reversal of the steady state with coupling $k_a$ is the same as the natural steady state with coupling $-k_a$~\cite{o2024geometric}.

\subsection{Model definitions}
\label{sec:models-methods}
We consider several models which are all consistent with Eqs.~(\ref{model_methods},\ref{eq:Fi}).  The number of particles is $M$.  In some cases it is useful to consider the particles as vertices of a polygon whose signed area is
\begin{equation}
A =\sum_{i=1}^M
 \frac{x_i y_{i+1}-y_i x_{i+1}}{2} .
 \label{eq:signed_A}
\end{equation}
in which indices are added modulo $M$, so $(x_{M+1},y_{M+1})=(x_1,y_1)$.

\paragraph{Two particle model.}  The model of Fig.~\ref{Fig1}(c) consists of two particles ($M=2$), connected by a single odd spring.  We write $\bm{r}_\mathrm{CM}=(\bm{r}_1+\bm{r}_2)/2$ for the position of the center of mass and write the interparticle vector in polar co-ordinates as $(\bm{r}_1-\bm{r}_2)=(r\cos\theta,r\sin\theta)$ so the energy becomes $E=\frac{k}{2}(r-r_0)^2$.  

\paragraph{Odd triangle model.}  Consider $M=3$ particles, with 3 odd springs connecting them.  
There are two distinct zero energy configurations which are equilateral triangles with $A=\pm \frac{\sqrt{3}}{4}$.

\paragraph{Odd square model.} Consider $M=4$ particles, with 4 identical odd springs connecting them as in Fig.~\ref{Fig2}.  For the quadrilateral with vertices $\bm{r}_1,\bm{r}_2,\bm{r}_3,\bm{r}_4$, we write $\psi_1,\psi_2,\psi_3,\psi_4$ for the four internal angles.  The square configuration has zero energy, as do rhombus configurations (corresponding to $\psi_1=\psi_3=\pi-\psi_2=\pi-\psi_4$).

We also consider a variant of this model which includes the angular potential $V_{\rm angle}$  of Eq.~\eqref{eq:V-angle}.  The resulting energy is
\begin{equation}
E = \sum_{i=1}^4 \left[ \frac{k}{2}(\norm{\bm{r}_{i+1}-\bm{r}_i} -r_0)^2  - \frac{\varepsilon}{4} \cos (2\psi_i) 
\right] 
\label{eq:square-energy}
\end{equation}
whose zero-energy configurations are the unfolded square and folded line of length $r_0$ or $2r_0$.

\paragraph{Double square network.}  We consider $M=6$ particles connected as in Fig.~\ref{Fig3}.  That is, the pairs of connected particles are $(1,2)$, $(2,3)$, $(3,4)$, $(4,5)$, $(5,6)$, $(6,1)$ and $(2,5)$.  Other pairs of particles interact via a Morse potential
\begin{equation}
    V_\mathrm{M}(r) = U_{\rm M} \left  (1-e^{-\chi_{\rm M}( r-r_0)}  \right )^2,
    \label{Morse potential}
\end{equation}
where $U_{\rm M}$ is the interaction strength, $r_0$ is the separation at which the interaction energy is minimal (which we take equal to the natural length of the harmonic springs), and $\chi_{\rm M} $ controls the width of the potential and is proportional to the square root of the curvature at the minimum.
We take representative parameters $U_{\rm M}=5 T$ and $\chi_{\rm M} = 5 / r_0$ throughout this work.

For the simulations of self-assembly with cargo [Fig.~\ref{Fig3}\textbf{d}–\textbf{e}], this double-square system was initialised in a zero-energy rectangular configuration.  It is assumed that on first reaching the ``chevron'' state, the cargo binds immediately and irreversibly, after which the system remains in the chevron state for all future times.  The definition of the chevron is based on the internal angles of the squares, details are given in Methods~\ref{sec:methods-numerics}.

\subsection{Analysis of two-particle model}
\label{Two-particle model (arbitrary $T$)}
As explained in Sec.~\ref{sec:models-methods}, the two-particle model is described in polar co-ordinates (detailed calculations are given in Supplementary Information).  The dynamics of the particle separation decouples from the centre of mass; converting the associated Langevin dynamics for $(r(t),\theta(t))$ to a Fokker-Planck equation for the probability density $P(r,\theta,t)$ yields
\begin{multline}
\frac{\partial P}{\partial t} =  \frac{\partial}{\partial r} \left[ 2\left( k(r-r_0) - \frac{T}{r} \right) P \right] 
-  \frac{2k_a(r-r_0)}{r} \frac{\partial P}{\partial \theta} 
\\+ 2T  \frac{\partial^2 P}{\partial r^2} + 
\frac{2T}{r^2}\frac{\partial^2 P}{\partial \theta^2}
\label{eq:fp-2particles}
\end{multline}
[see also Eq.~\eqref{eq:fp-curvi} below, for the general form of Fokker-Planck equations in curvilinear co-ordinates].  
One may verify that the associated  stationary distribution is
\begin{equation}
P^\star(r,\theta) = C r \exp \left[-\frac{k}{2T}(r-r_0)^2\right],
\label{two particles proba}
\end{equation}
where $C=C(T,k,r_0)$ is a normalization constant. 
As already shown in Sec.~\ref{Langevin}, this distribution is independent of $k_a$.

The steady-state probability current can be read from Eqs.~(\ref{eq:fp-2particles},\ref{two particles proba}) as
\begin{equation}
\bm{j}_\mathrm{p}^\star(r,\theta) =\frac{2 k_a(r-r_0)P^\star(r,\theta)}{r} {\bm{e}}_{\theta},
\label{two particles J}
\end{equation}
where ${\bm{e}}_{\theta}$ is a unit vector in the $\theta$ direction.  Hence the current corresponds to a steady rotation of the particles about their center of mass.
The associated mean angular velocity is $\langle\dot\theta\rangle=\int (\bm{j}_\mathrm{p}^\star \cdot \bm{e}_\theta) \mathrm{d} r \mathrm{d}\theta$ which is non-zero in general, it is exponentially suppressed at low temperatures ($T \ll k r_0^2$) as
\begin{equation}
\langle \dot \theta \rangle \approx k_a \sqrt{\frac{2T}{\pi kr_0^2}}  \exp\left(-\frac{kr_0^2}{2T} \right) \; .
\end{equation}

\subsection{Dynamics of the odd square}
\label{Derivation of the reduced beta dynamics}

For the odd square model, we explain in the main text that the low-temperature dynamics of the internal angle $\beta$ is described by Eq.~\eqref{beta dynamics}.  We outline the derivation of this result, with details in Supplementary Information.  The central assumption is that the configuration remains always close to a rhombus shape, with small deviations that we express in terms of the four finite-frequency normal modes of the associated spring network.  Throughout this section we set $r_0=1$, this fixes the units of length but does not lose any generality.

\paragraph{Normal modes and small deformations.}
As in Sec.~\ref{Langevin}, we write
$\bm{X}=(x_1,y_1,\dots,x_4,y_4)$. We parameterise this (see Supplementary Information for full details) as:
\begin{equation}
\bm{X} = \bm{X}_{\rm CM} + \Matrix{R}(\theta)\Big(\bm{s}^0(\beta)+\sum_{k\in{\cal M}}\alpha_k\bm{z}_k(\beta)\Big),
\end{equation}
where $\bm{X}_{\rm CM}$ accounts for the centre of mass motion, $\Matrix{R}(\theta)$ is an $8\times 8$ orthogonal matrix that accounts for rigid body rotations, $\bm{s}^0$ is the reference configuration of the rhombus (with internal angle $\beta$) and $\bm{z}_k(\beta)$ is a normal mode, indexed by $k\in {\cal M}$ with ${\cal M}=\{c,d,e,f\}$ labelling the modes.  The small amplitudes of these normal modes are the $\alpha_k$'s so the network configuration is described by a set of curvilinear co-ordinates
\begin{equation}
\bm{q}=(x_{\rm CM},y_{\rm CM},\theta,\beta,\alpha_c,\alpha_d,\alpha_e,\alpha_f)
\end{equation}
(In addition to its free rotation described by $\theta$, the rhombus has a soft mode which is accounted for by the angle $\beta$.)  

We consider small deformations  $\alpha_k \ll 1$ and weak angular interaction $\varepsilon\ll k$.  In this case the energy from Eq.~\eqref{eq:square-energy} can be approximated as  
\begin{multline}
E  \approx k\left[\alpha_c^2+\alpha_d^2+(1-\cos\beta)\alpha_e^2+(1+\cos\beta)\alpha_f^2\right] \\
 -\frac{1}{2} \varepsilon \left(1+\cos(4 \beta)\right).
\end{multline}
For fixed $\beta\neq 0,\pi$, the associated Boltzmann distribution has $\alpha = O(\sqrt{T/k})$ so the assumption of small deformation is valid at low temperature.  In the following we also assume $\varepsilon=O(T)$ so that the two contributions to the energy are of the same order.
The Jacobian $\Matrix{J}$ of the transformation 
$(\bm{X}\mapsto\bm{q})$ has elements $J_{nm}=\partial X_n/\partial q_m$ and for small deformations we also find  
$$
\det(\Matrix{J}) \approx \frac{1}{2}\left(2-\alpha_c^2-2\sqrt{2}\,\alpha_d+\alpha_d^2\right) \; .
$$

\paragraph{Fokker--Planck equation and separation of time scales.}
We construct the Fokker-Planck equation associated with the system's Langevin dynamics, in curvilinear coordinates.  This involves the metric ${g}=\Matrix{J}^T \Matrix{J}$.  The probability density is for $\bm{q}$ is $P=P(\bm{q},t)$ and we have in general~\cite{risken1989fokker}
\begin{align}
\begin{split}
    &\frac{\partial P}{\partial t} = \sum_{nm} \frac{\partial }{\partial q_n} \left [ (g^{-1})_{nm} \left [  V  \frac{\partial }{\partial q_m} \left( \frac{T}{V} P \right) - \mu_m P \right ]\right ]
\end{split}
\label{eq:fp-curvi}
\end{align}
where $\mu_m= \sum_{l=1}^8 F_l (\bm{X}) J_{lm}$ is a generalised force and $V=\sqrt{\det g}$.

In the low-temperature regime then the dynamics of the $\alpha_k$ are fast compared to $\beta$ and $\theta$.   Our aim is to analyse the dynamics of $\beta$, which is independent of $\theta$ by rotational invariance.  Hence $\theta$ plays no role in the following.  The separation of time scales can then be exploited to derive the dynamics of $\beta$, following~\cite{pavliotis2008multiscale}. The calculation proceeds in two stages.   

First, the variables $(\alpha_d,\alpha_e,\alpha_f)$ are eliminated.  Averaging the fast dynamics at fixed $\beta$ one finds
    \begin{equation}
  \left\langle \alpha_f^2-\alpha_e^2\right\rangle_{\beta} = -\frac{T \cos \beta}{k \sin^2\beta } 
      \label{eq:avg_af2_me_ae2}
    \end{equation}
which shows that the fluctuations in $\alpha_f$ and $\alpha_e$ depend strongly on $\beta$.  In particular, for $\beta=0,\pi$, the fluctuations can diverge, signalling a breakdown of the assumption of small deformation.  The entropy associated with these fluctuations means that such values of $\beta$ have larger statistical weight in equilibrium, leading to a generalised force $ k \sin\beta \langle \alpha_f^2-\alpha_e^2\rangle_{\beta}$ in the joint dynamics of $\beta,\alpha_c$, so that
\begin{align}
\begin{split}
\begin{pmatrix}
        \dot{\beta} \\
        \dot{\alpha_c} \\
\end{pmatrix}
=
&
\begin{pmatrix}
        -2 T \cot \beta \\
        -2k \alpha_c  \\
\end{pmatrix}
    + 
\begin{pmatrix}
        -2 \sqrt{2} \alpha_c k_a  \\
         \frac{\sqrt{2} T k_a \cot \beta}{k} \\
\end{pmatrix}
+
\begin{pmatrix}
        2\sqrt{T } \eta^{\beta} \\
        \sqrt{2T} \eta^{\alpha_c} \\
\end{pmatrix}.
\label{Beta and alphac dynamics approx}
\end{split}
\end{align}

Second, the variable $\alpha_c$ is integrated out: the separation of time scales allows~\cite{pavliotis2008multiscale} adiabatic elimination of this variable, which amounts to setting $\dot{\alpha}_c = 0 $ to yield
\begin{equation}
    \alpha_c = \frac{\sqrt{2} T k_a \cot \beta}{2k^2} + \sqrt{\frac{T}{2k^2}} \eta^{\alpha_c}. 
\end{equation}
Substituting this back into the $\beta$ dynamics, the  result is: 
\begin{equation}
\dot\beta = -2T \Lambda\cot{\beta}-4 \varepsilon \Lambda \sin (4 \beta) + 2\sqrt{T \Lambda}\, \eta,
\label{eq:beta_sde_methods}
\end{equation}
with 
$
\Lambda = 1 + (k_a/k)^2
$ as in Eq.~\eqref{eq:Lambda}.
For $\varepsilon=0$ then  Eq.~\eqref{eq:beta_sde_methods} reduces to Eq.~\eqref{beta dynamics}.

\paragraph{Remarks and limitations.}
Equation~\eqref{eq:beta_sde_methods} relies on the low-temperature expansion, a clear separation of timescales between the fast shape modes $(\alpha_d,\alpha_e,\alpha_f)$ and the slow variables $(\beta,\alpha_c)$.  Near $\beta \to 0,\pi$ the $\cot\beta$ term becomes large and higher-order corrections are significant~\cite{rocklin2018folding}.

\subsection{Numerical simulations}
\label{sec:methods-numerics}

\paragraph{Equations of motion.}
All stochastic dynamics were integrated using an Euler-Maruyama scheme. The particle positions were updated according to:
\begin{equation}
\bm{r}_i(t+\Delta t) = \bm{r}_i(t) + \bm{F}_i(t)\Delta t + \sqrt{2T\Delta t}\,\bm{\xi}_i(t),
\end{equation}
with $\bm{F}_i(t)$ as in Eq.~\eqref{eq:Fi}, the time step is $\Delta t$  and the components of the noise vectors $\bm{\xi}_i$ were sampled as independent standard normal variates and truncated to the interval $[-\xi_{\mathrm{max}},\xi_{\mathrm{max}}]$. In simulations, $\xi_{\mathrm{max}}=4$ was used. The truncation suppresses rare large noise increments that would otherwise lead to numerical instabilities at finite time step (see SUPP).

\paragraph{Angular velocity computation for the triangle model}
In the case of the triangle model, the angular velocity as shown in Fig.~\ref{Fig1}\textbf{d} was extracted from particle motion in the centre-of-mass frame. Positions were measured relative to the centre of mass as $\bm{r}_{i, \text{cm}} = \bm{r}_i - \bm{r}_\mathrm{cm} $,
 and velocities ($\{ \bm{v}_i \}$) were obtained via central finite differences. The instantaneous angular momentum:
\begin{equation}
L(t) = \sum_i (\bm{r}_{i, \text{cm}} \times \bm{v}_i)_z ,
\end{equation}
and moment of inertia:
\begin{equation}
I(t) = \sum_i |\bm{r}_{i, \text{cm}}|^2 ,
\end{equation}
yield the angular velocity through $\omega(t) = L(t)/I(t)$. This definition isolates collective rotation while removing translational contributions.
\paragraph{Power computation.}
To quantify the energetic cost of activity in the four-particle square system, we first computed the total mechanical work performed on the system during a single assembly event. 

The work was evaluated using a Stratonovich discretization of the stochastic integral,
\begin{equation}
W = \sum_i \int \bm{F}_i \circ d{\bm{r}}_i
\end{equation}

where $\bm{F}_i$ denotes the total force acting on particle $i$.
 The average power input during assembly was then defined as the work per unit time:
\begin{equation}
P = \frac{W(\tau)}{\tau}.
\end{equation}
where $\tau$ is the assembly time (see Methods Sec.~\ref{Flip rates estimation}).

The average power $P$ was computed separately for each trajectory and then averaged over $N$ independent realizations. 

\subsection{Flip rates $R$ and first-passage times $\tau$ estimation}
\label{Flip rates estimation}

Flip rates and first-passage times are defined in terms of basins arounds target states, defined using thresholds on suitable order parameters, as appropriate for each model (see below).  For any given initial condition, the mean first passage time is obtained by measuring first time at which the system enters the target basin, averaged over many stochastic trajectories.

Flip rates between target states are defined within the steady state of the (stochastic) dynamics, which is reached by running the dynamics for a time $t_{\mathrm{start}}$ before any measurements are made. After this, transition times between states are measured, where states are defined as basins around the target configurations using thresholds on suitable order parameters (see below), and configurations outside both thresholds are not assigned to either state. 

For a system with two states $S_1,S_2$, the transition rate from $S_1$ to $S_2$ is defined as
\begin{equation}
R_{12} = \langle \tau_{1 \to 2} \rangle^{-1},
\label{eq:flip-rate-def}
\end{equation}
where $\langle \tau_{1 \to 2} \rangle$ is the mean time between an entry into the basin of $S_1$ and the next entry into the basin of $S_2$. During this interval the system may temporarily leave the basin of $S_1$, but must not enter the basin of $S_2$. The statistical uncertainty on $R_{12}$ is obtained using standard error propagation.

\paragraph{Odd triangle}
For the odd triangle system, we consider two target basins which are $A>A_c$ and $A<-A_c$, where $A$ is the signed area of Eq.~\eqref{eq:signed_A} and $A_c=\sqrt{3}/8$.
The assembled state includes configurations from both target basins; the assembly time $\tau$ is the mean first passage time to this assembled state, for a system initialised as a line, in which two springs have their equilibrium length $r_0$, and one is stretched to $2 r_0$.  

The flip rate $R$ shown Fig.~\ref{Fig2}\textbf{e} was measured as in Eq.~\eqref{eq:flip-rate-def}, as the transition rate between the two target basins with positive and negative $A$.

\paragraph{Square folding time $\tau$.}
Simulations are initialized in a square configuration with $|A| = 1$. The target basin for the folded state is $|A|<0.05$.

\paragraph{Double-square: flip rates between ladder and chevron.}
Figure~\ref{Fig3}\textbf{a} shows ladder and chevron configurations of the double-square system, as well as the  two internal angles $\beta', \beta''$ that are used to define their target basins.  We define the order parameter
$\bar{\beta} = \tfrac{1}{2} (\beta' + \beta'')$.
The ladder and chevron basins are centered at $\bar{\beta} = \pi$ and $\bar{\beta} = \beta_c = 2\arccos\!\left((2-r_0^2)/2\right)$, respectively. The target basin for each state is that $\bar{\beta}$ is within a threshold $\delta = \pi/8$ of the basin center and is closer to that center than to the other; the threshold lines are presented in Fig.~\ref{Fig3}.\textbf{a},\textbf{b}.  
The flip rate between these two states was measured as in Eq.~\eqref{eq:flip-rate-def}.
\subsection{Supplementary Video}
Supplementary Movies 1 and 2 provide representative stochastic trajectories illustrating the dynamics discussed in the main text. Movie 1 (based on Fig.\ref{Fig2}\textbf{b}) shows folding of the four-particle square network from a square to a line state under passive ($k_a=0$) and non-reciprocal ($k_a=1$) dynamics, together with the time evolution of the signed area used to quantify folding. Movie 2 (based on Fig.\ref{Fig3}\textbf{a}) shows trajectories of the six-particle double-square network for both $k_a=0$ and $k_a=1$, with configurations accompanied by the time evolution of the mean internal angle distinguishing ladder and chevron states. In both cases, non-reciprocal forces accelerate the dynamics, enhancing folding in the four-particle system and increasing transition rates between metastable states in the six-particle system.

\newpage
\onecolumngrid
\section*{Supplementary Information}
\noindent In this supplementary material, we provide a detailed derivation of the low-temperature stochastic dynamics for the various models of odd mechanical networks of overdamped thermal colloids introduced in the main text. 
Sec.~\ref{sm:spring} is dedicated to the odd-spring model. Here we introduce the Fokker-Planck equation in curvilinear coordinate and derive the stationary distribution and angular current.
Sec.~\ref{sm:triangle} deals with the triangle model, for which we demonstrate the emergence of couplings between the equilibrium elastic eigenmodes mediated by odd forces.
A more involved version of the same derivation is presented in Sec.~\ref{sm:square} for the square model, where we additionally apply adiabatic elimination to project the full dynamics onto the rhombus folding pathway, thus obtaining an explicit expression for the activity-induced speedup factor.
Finally, in Sec.~\ref{sm:numerics} we give a detailed description of the methods used in numerical experiments. 

\subsection*{Two particle odd-spring model}\label{sm:spring}
\indent We consider a two-particle system as defined in Methods \textbf{A.2}, deriving the equations presented in Methods \textbf{A.3}. For this specific system, the dynamics can be solved analytically by decomposing them into the dynamics of the centre of mass $\bm{r}_\mathrm{CM}={(\bm{r}_1+\bm{r}_2)/2}$, which are purely diffusive, and of the relative distance between two particles $\bm{r}=\bm{r}_1-\bm{r}_2$:
\begin{align}
\begin{split}
    & \dot{\bm{r}}_\mathrm{CM} = \sqrt{T} \bm{\eta}_\mathrm{CM}, \\
    & \dot{\bm{r}}= -2\left (\Matrix{I}+\frac{k_a}{k} \Matrix{\varepsilon} \right )  \nabla_{\bm{r}} E  +\sqrt{4T} \bm{\eta},
\end{split}
\end{align}
where all noise components are Gaussian white and independent. The interaction energy $E$ is given in the main text as $E=k(|\bm{r}|-r_0)^2/2$.  To characterise the stationary state, we perform a change of variables from cartesian $\bm{X} = (x_1, y_1, x_2, y_2)$ to cylindrical $\bm{q}=\{x_\mathrm{CM}, y_\mathrm{CM}, r, \theta \}$ coordinates, with $r , \theta$ defined by:
\begin{align}
\begin{split}
    & x_1 = x_\mathrm{CM} + \frac{r}{2} \cos \theta, \\
    & y_1 = y_\mathrm{CM} + \frac{r}{2} \sin \theta, \\
    & x_2 = x_\mathrm{CM} - \frac{r}{2} \cos \theta, \\
    & y_2 = y_\mathrm{CM} - \frac{r}{2} \sin \theta . \\
\end{split}
\end{align}
The Jacobian for this transformation is $J_{im}=\partial X_i/\partial q_m$, specifically
\begin{equation}
    \Matrix{J} = \left(
\begin{array}{cccc}
 1 & 0 & \frac{\cos (\theta )}{2} & -\frac{1}{2} r \sin (\theta ) \\
 0 & 1 & \frac{\sin (\theta )}{2} & \frac{1}{2} r \cos (\theta ) \\
 1 & 0 & -\frac{\cos (\theta )}{2} & \frac{1}{2} r \sin (\theta ) \\
 0 & 1 & -\frac{\sin (\theta )}{2} & -\frac{1}{2} r \cos (\theta ) \\
\end{array}
\right),
\label{eq:rpol}
\end{equation}

Then the curvilinear form of the Fokker-Planck equation~\cite{risken1989fokker} is
\begin{align}
\begin{split}
    \frac{\partial P(\bm{q} ;t)}{\partial t} = 
    &\sum_{n,m} \frac{\partial^2 }{\partial q_n \partial q_m} (T g^{nm} P(\bm{q} ;t)) -\sum_{n,m} \frac{\partial }{\partial q_n} \left [ \left (g^{nm} \mu_m + T \left(\frac{\partial }{\partial q_m} g^{nm}\right ) + \frac{1}{2}T   g^{nm} \frac{\partial }{\partial q_m} \log (\det g) \right )    P(\bm{q} ;t)  \right ].
\end{split}
\label{suppeq:fp-curvi}
\end{align}
where
the metric ${g}=\Matrix{J}^T \Matrix{J}$ and $g^{nm} = (g^{-1})_{nm}$,  also $\mu_m= \sum_i F_i (\bm{X}) J_{im}$ is a generalized force.   
Substituting the co-ordinates and simplifying, the centre-of-mass motion decouples and one arrives at an equation for the probability distribution of the relative co-ordinates alone:
\begin{equation}
\frac{\partial P(r,\theta ;t)}{\partial t} 
=  \frac{\partial}{\partial r} \left[ 2\left( k(r-r_0) - \frac{T}{r} \right) P(r,\theta ;t) \right] 
-  \frac{2k_a(r-r_0)}{r} \frac{\partial P(r,\theta ;t)}{\partial \theta} 
+ 2T  \frac{\partial^2 P(r,\theta ;t)}{\partial r^2} + 
\frac{2T}{r^2}\frac{\partial^2 P(r,\theta ;t)}{\partial \theta^2}
\label{eq:fp-2particle}
\end{equation}
as in the main text.
%

This Fokker-Planck equation can additionally be expressed in terms of the probability current $\bm{j}_{\rm p}$ as
 \begin{equation}
     \frac{\partial P(r,\theta ;t)}{\partial t}  = -\frac{\partial}{\partial r} \left[ ({\bm{j}_{\rm p}(r,\theta ;t) \cdot \bm{e}_r})r\right]
     - 
     \frac{\partial}{\partial \theta} (\bm{j}_{\rm p}(r,\theta ;t)\cdot\bm{e}_\theta )
     \label{eq:fp-2part-supp}
 \end{equation}
with
\begin{equation}
    \bm{j}_\mathrm{p} (r,\theta; t) = \frac{2(r-r_0)}{r}(-k\bm{e}_r + k_a \bm{e}_\theta ) P(r,\theta; t) - 2T \left[ 
      \frac{\partial}{\partial r}   \left(\frac{ P(r,\theta; t)}{r} \right)  \bm{e}_r+
       \frac{1}{r^2} \frac{\partial }{\partial \theta} 
     P(r,\theta; t) \bm{e}_\theta
     \right] .
\end{equation}
At steady-state, probability conservation imposes $\bm{j}_\mathrm{p} (\bm{r}; t)\cdot \bm{e}_r=0$ so that the stationary current is purely rotational, as given
in the main text.

As a final comment, we note that the form Eq.~\eqref{eq:fp-2part-supp} makes it clear that the probability density $P$ remains normalised under time evolution, that is $\frac{\partial}{\partial t} \int P(r,\theta;t) dr d\theta = 0$.  However, it is also possible to work with a covariant density $p(r,\theta;t)=P(r,\theta;t)/r$ whose normalisation condition is $\frac{\partial}{\partial t} \int p(r,\theta;t) r dr d\theta = 0$, where we recognise the standard integration measure for plane polar co-ordinates.  In this case the Fokker-Planck equation becomes
 \begin{equation}
     \frac{\partial p}{\partial t}  = -\frac{1}{r} \frac{\partial}{\partial r} \left[ ({\bm{j}_{\rm p} \cdot \bm{e}_r})r\right]
     - 
     \frac{1}{r} \frac{\partial}{\partial \theta} (\bm{j}_{\rm p}\cdot\bm{e}_\theta )
 \end{equation}
 and we recognise the right hand side as $(-\operatorname{div}\bm{j}_{\rm p})$, expressed in plane polar co-ordinates.  This confirms that $\bm{j}_{\rm p}$ is  the covariant probability current.

\subsection*{Triangle model: linearised mode dynamics and flip mechanism} \label{sm:triangle}
\label{SI:triangle_modes}
We analyse the dynamics of the three-particle triangle system (as defined in Methods \textbf{A.2}) in the low-temperature regime, $T \ll k r_0^2$, focusing on the onset of spontaneous rotations and enhancements of chirality-flip rates in the presence of odd forces. In this limit, thermal fluctuations are weak and the system configuration remain close to the mechanical equilibrium configuration at all times, allowing the nonlinear stochastic dynamics to be approximated by a quadratic expansion of the energy around equilibrium. Without loss of generality, we henceforth set our units of length such that $r_0 = 1$.

\begin{figure*}[t]
    \centering
    \includegraphics[]{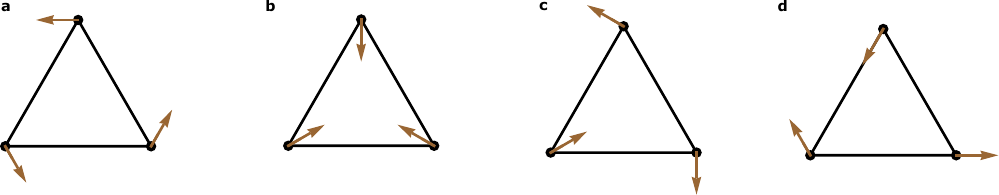}
    \caption{\textbf{Elastic eigenmodes of a triangle.} \textbf{a}, Anticlockwise rotation ($z_1$). \textbf{b}, Compression ($z_2$). \textbf{c}, Shear I ($z_3$), which is responsible for the triangle flips. \textbf{d}, Shear II ($z_4$).}
    \label{Triangle modes}
\end{figure*}

We now introduce a suitable set of curvilinear co-ordinates to describe this model.  As an equilibrium reference configuration of the triangle, we take
$\bm{x}_1^0=(x^0_1,y_1^0)=(-0.5,0)$, $\bm{x}_2^0=(x_2^0,y_2^0)=(0, \frac{\sqrt{3}}{2})$, $\bm{x}_3^0=(x_3^0,y_3^0)=(0.5,0)$ and $\bm{X}^0=(\bm{x}_1^0,\bm{x}_2^0,\bm{x}_3^0)$.
Considering small deformations of the (passive) spring network about this configuration, the (normalised)  eigenmodes are
\begin{align}
\begin{split}
    &  \bm{\xi}_x =\frac{1}{\sqrt{3}} \left (1,0,1,0,1,0\right ), \\
    & \bm{\xi}_y =\frac{1}{\sqrt{3}} \left (0,1,0,1,0,1\right ), \\
    & \bm{z}_1=\frac{1}{2\sqrt{3}} \left (1, -\sqrt{3}, -2, 0, 1, \sqrt{3} \right) , \\
    & \bm{z}_2= \frac{1}{2 \sqrt{3}} \left (\sqrt{3},1,0,-2,-\sqrt{3},1 \right), \\
    & \bm{z}_3= \frac{1}{2\sqrt{3}}\left (\sqrt{3},1,-\sqrt{3},1,0,-2\right ),
     \\
    & \bm{z}_4=  \frac{1}{2 \sqrt{3}}\left (-1,\sqrt{3},-1,-\sqrt{3},2,0\right ).
    \label{passive basis triangle}
\end{split}
\end{align}
which correspond to two pure translations $\bm{\xi}_{1,2}$, a free rotation ($\bm{z}_1$) and three non-trivial deformation modes ($\bm{z}_2,\bm{z}_3,\bm{z}_4$), see Fig.~\ref{Triangle modes}.
Then we write a general configuration of the triangle as 
\begin{equation}
    \bm{X} =  (x_{\rm CM} \bm{\xi}_x + y_{\rm CM} \bm{\xi}_y)\sqrt{3} + \Matrix{R}(\theta)\Big(\bm{X}^0+\sum_{k=2}^4 \alpha_k\bm{z}_k(\beta)\Big),
    \label{decomposition-triag}
\end{equation}
where the centre-of-mass position is $(x_{\rm CM},y_{\rm CM})$, also $\Matrix{R}(\theta)$ describes a rigid body rotation by angle $\theta$; the coefficients $\alpha_2,\alpha_3,\alpha_4$ describe the deformation of the triangle.  There is no term like $\alpha_1\bm{z}_1$ in Eq.~\eqref{decomposition-triag} because the rotational motion is accounted for instead by $\theta$.
The resulting co-ordinates are then $$(x_{\rm CM},y_{\rm CM},\theta,\alpha_2,\alpha_3,\alpha_4) .$$

Small deformations of the triangle correspond to small $\alpha$. 
Expanding the total elastic energy up to second order in these parameters results in a linear dynamics, which can be written compactly as a vector Langevin equation,
\begin{equation}
    \dot{\bm{\alpha}} = -\Matrix{A}\cdot\bm{\alpha} + \sqrt{2T} \bm{\eta},
    \label{eq:alphadot}
\end{equation}
%
%
%
with
\begin{equation}
\bm{\alpha} = \left( \begin{array}{c} \alpha_2 \\ \alpha_3 \\ \alpha_4 \end{array} \right), \qquad
    \Matrix{A}=\left(
\begin{array}{cccccc}
 3 k & 0 & 0 \\
 0 & \frac{3 k}{2} &\frac{3 k_a}{2} \\
 0 & -\frac{3 k_a}{2} & \frac{3 k}{2} \\
\end{array}
\right),
\label{A matrix passive basis}
\end{equation}
The rotational motion is additionally described (for small deformations) as
\begin{equation}
\dot\theta = -3k_a \alpha_2
\label{eq:dot-theta-triag}
\end{equation}

The off-diagonal elements of $\Matrix{A}$ are proportional to $k_a$, they represent activity-induced couplings between otherwise independent passive (shear) modes.  This coupling enhances excitation of $\bm{z}_3$, in turn leading to more frequent triangle flips.
Similarly Eq.~\eqref{eq:dot-theta-triag} couples dilation of the triangle to rotation.
In particular, the compression mode ($\bm{z}_2$) drives rotations ($\bm{z}_1$), while the two shear modes form an activity-coupled pair.
These activity-induced couplings mimic the stress–strain relations derived in continuum theories of odd elasticity~\cite{scheibner2020odd}, in that shear modes couple with each other while dilational stress leads to rotational strain.

Note that Eq.~\eqref{eq:alphadot} is an approximation for small deformations and low temperatures; it neglects additional terms at $O(T)$ and $O(\alpha^2)$.  (These are derived below for the case of the square network.)  The main qualitative effect of these terms for the triangle is that $\alpha_2$ acquires a negative average value in the steady state, corresponding to dilation of the triangle, due to the increased entropy on expansion.  (A similar effect was already noted for the single spring, analysed above.)  
Using Eq.~\eqref{eq:dot-theta-triag} and that $\langle \alpha_2 \rangle <0$, the non-reciprocal coupling results in spontaneous rotations with chirality controlled by the sign of $k_a$.


\subsection*{Square model: linearisation and adiabatic elimination} \label{sm:square}
We now consider the four-particle network defined in Methods~\textbf{A.2}, and analize the effective dynamics of the internal angle $\beta$. Here, $\beta$ parametrises a continuous family of near-zero–energy configurations (floppy mode) in which the square can deform into a rhombus and, in the limiting cases $\beta \to 0$ or $\beta \to  \pi$, collapse into a line configuration.  Throughout the derivation, we work in the low-temperature regime, $T \ll k r_0^2$, such that thermal fluctuations perturb the passive mechanical equilibrium configuration only weakly. In addition, we assume that the angular interactions, as defined in Eq.~(4) of the main text, are weak, with $\varepsilon \sim T$. This hierarchy of energy scales allows for a controlled time scale separation between fast vibrational modes and slow folding dynamics. Under these assumptions, the full multidimensional stochastic dynamics can be systematically reduced to an effective stochastic equation governing the evolution of the internal angle $\beta$.


\subsubsection{Passive eigenmodes and curvilinear coordinates}

Consistent with our low-$T$ assumption, we consider configurations for which distortions relative to the one-parameter family of passive mechanical equilibrium configurations along the quasi-floppy mode are small. This corresponds to deformations that preserve bond lengths to leading order but allow the internal angle $\beta$ to vary.

\begin{figure*}[t]
    \centering
    \includegraphics[]{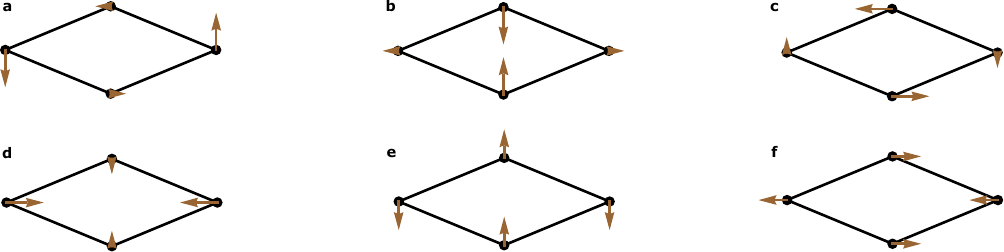}
    \caption{\textbf{Eigenmodes of a rhombus.} \textbf{a}, Rotation ($z_a$). \textbf{b}, Flattening of a rhombus ($z_b$), which is the floppy mode causing the folding. \textbf{c}, Antisymmetric shearing ($z_c$). \textbf{d}, Contraction ($z_d$). \textbf{e}, Symmetric shearing I ($z_e$). \textbf{f}, Symmetric shearing II ($z_f$).}
    \label{rhombus eigenmodes}
\end{figure*}

To this end, we introduce appropriate curvilinear co-ordinates, similar to Eq.~\eqref{decomposition-triag}. 
%
The equilibrium reference configuration now depends explicitly on the internal angle $\beta$, as defined in Fig. 2a of the main text:
\begin{align}
\begin{split}
&(x_1^0,y_1^0)= \left ( \cos \frac{\beta}{2}, 0  \right ), \\
&(x_2^0,y_2^0)= \left ( 0, \sin \frac{\beta}{2} \right ), \\
&(x_3^0,y_3^0)= \left ( -\cos \frac{\beta}{2}, 0  \right ), \\
&(x_4^0,y_4^0)= \left (0, -\sin \frac{\beta}{2} \right ).
\end{split}
\end{align}
and we write
\begin{equation}
    \bm{s}^0 = \left ( x_1^0,y_1^0 , x_2^0,y_2^0, x_3^0,y_3^0, x_4^0,y_4^0 \right ) 
\end{equation}
Considering small deformations about this state, the eigenmodes of the square spring network are
%
%
\begin{align}
\begin{split}
\bm{\xi}_x 
&= \frac{1}{2}\left(1,0,1,0,1,0,1,0\right), \\
\bm{\xi}_y 
&= \frac{1}{2}\left(0,1,0,1,0,1,0,1\right), \\
\bm{z}_a &= \frac{1}{\sqrt{2}}\left( 0, \cos\left(\frac{\beta}{2}\right), -\sin\left(\frac{\beta}{2}\right), 0, 0, -\cos\left(\frac{\beta}{2}\right), \sin\left(\frac{\beta}{2}\right), 0 \right), \\
\bm{z}_b &= \frac{1}{\sqrt{2}}\left( \sin\left(\frac{\beta}{2}\right), 0, 0, -\cos\left(\frac{\beta}{2}\right), -\sin\left(\frac{\beta}{2}\right), 0, 0, \cos\left(\frac{\beta}{2}\right) \right), \\
\bm{z}_c &= \frac{1}{\sqrt{2}}\left( 0, -\sin\left(\frac{\beta}{2}\right), -\cos\left(\frac{\beta}{2}\right), 0, 0, \sin\left(\frac{\beta}{2}\right), \cos\left(\frac{\beta}{2}\right), 0 \right), \\
\bm{z}_d &= \frac{1}{\sqrt{2}}\left( -\cos\left(\frac{\beta}{2}\right), 0, 0, -\sin\left(\frac{\beta}{2}\right), \cos\left(\frac{\beta}{2}\right), 0, 0, \sin\left(\frac{\beta}{2}\right) \right), \\
\bm{z}_e &= \frac{1}{2}\left( 0, -1, 0, 1, 0, -1, 0, 1 \right), \\
\bm{z}_f &= \frac{1}{2}\left( -1, 0, 1, 0, -1, 0, 1, 0 \right).
\end{split}
\label{rhombus beta basis}
\end{align}
See Fig.~\ref{rhombus eigenmodes}.  Then
similar to Eq.~\eqref{decomposition-triag}, we write
\begin{equation}
    \bm{X} = \bm{X}_{\rm CM} + \Matrix{R}(\theta)\Big(\bm{s}^0(\beta)+\sum_{k\in{\cal M}}\alpha_k\bm{z}_k(\beta)\Big),
    \label{decomposition}
\end{equation}
with $\bm{X}_{\rm CM} = 2(x_{\rm CM}\bm{\xi}_x + y_{\rm CM}\bm{\xi}_y)$ accounting for the centre of mass, and ${\cal M}=\{c,d,e,f\}$.

The modes $\bm{z}_{a,b}$ do not appear in \eqref{decomposition} because they correspond to the free rotation mode ($a$, which is accounted or by the co-ordinate $\theta$) and the soft deformation mode ($b$, which is accounted for by $\beta$).
As we will show later, entropic effects favour deformations along the soft ($b$) mode, leading to a bias towards collapse of the square into a line configuration through rhombus-like intermediates. Because the resulting variation of $\beta$ is not small in general, no individual fixed-$\beta$ eigenmode in~\eqref{rhombus beta basis} cannot serve as a reference for the folding; instead, the reference configuration must evolve continuously with $\beta$.
The full set of curvilinear co-ordinates is then taken as
$$
\bm{q}=(2x_{\rm CM},2y_{\rm CM},\theta,\beta,\alpha_c,\alpha_d,\alpha_e,\alpha_f)
$$
(The factors of $2$ multiplying the centre of mass co-ordinates are arbitrary; they are convenient for the following discussion.)
%
%
The change of variables from Cartesian to curvilinear coordinates introduces a nontrivial Jacobian, whose  determinant is 
\begin{equation}
    |\Matrix{J}|=\frac{1}{2}\left|2-\alpha_c^2 - 2\sqrt{2} \alpha_d + \alpha_d^2\right| \approx 
       1
    - \sqrt{2} \alpha_d 
   + O(\alpha^2) 
    ~, \label{eq:jacob_square}
\end{equation}
where the second equality relies on the assumption that fluctuations of the fast elastic modes are small, $\alpha \ll 1$, whereby the absolute value can be dropped.

\subsubsection{Linearisation of fast modes}

In order to derive effective dynamics within the small-$\alpha$ approximation, we first expand the total elastic potential to quadratic order in the fast eigenmode amplitudes. This gives
\begin{equation}\label{eq:eff_V_spring}
    V_{\rm spring} (\bm{q})\approx k \left[\alpha_c^2 + \alpha_d^2 + (1  - \cos{\beta}) \alpha_e^2 + (1  + \cos{\beta}) \alpha_f^2\right]~.
\end{equation}
For fixed $\beta$, the characteristic amplitudes are thus $O(\sqrt{T})$.
Next, we express the angle potential in the curvilinear coordinates $\bm{q}$. For the internal angle $\psi_i$ corresponding to the vertex occupied by the $i$th particle, one has
\begin{equation}
    \psi_i (\bm{X}) = \arccos \left (\frac{\bm{r}_{i,i-1}\cdot\bm{r}_{i+1,i}}{|\bm{r}_{i,i-1}||\bm{r}_{i+1,i}|}  \right ),
\end{equation}
where $\bm{r}_{i,k} = \bm{r}_{i}- \bm{r}_{k}$ and indices are computed modulo 4. Together with the identity $\cos^2 (2 \arccos (x))=(2x^2-1)^2$, the above yields the alternative form for Eq.~(4):
\begin{equation}
    V_{\rm angle} (\bm{X})= -\frac{\varepsilon}{4}\sum_{i=1}^4  \left[ 2\left (\frac{\bm{r}_{i,i-1}\cdot\bm{r}_{i+1,i}}{|\bm{r}_{i,i-1}||\bm{r}_{i+1,i}|}  \right )^2-1\right]^2.
\end{equation}
Rewriting this in curvilinear variables and expanding to second order in $\alpha_k$ gives
\begin{equation}
    V_{\rm angle} (\bm{q})   = \frac{\varepsilon}{4}  \left[\left(\alpha_f^2-\alpha_e^2\right) \cos (3 \beta)+\left(-4 \alpha_e^2-4 \alpha_f^2+2\right) \cos (4 \beta)+3 \left(\alpha_f^2-\alpha_e^2\right) \cos (5 \beta)+2\right].
\end{equation}
Further assuming that $\varepsilon \sim T$ and retaining only the terms of lowest order in small $T$, gives the simple expression:
\begin{equation}\label{eq:eff_V_angle}
    V_{\rm angle} (\bm{q})  \approx -\frac{\varepsilon}{2}(1+\cos (4 \beta)) .
\end{equation}

\subsubsection{Dynamics in curvilinear co-ordinates}

In order to derive an effective equation for the folding angle $\beta$, all of the other variables need to be integrated out.  We first write the Fokker-Planck equation in the appropriate co-ordinate system, under our assumption of low temperature. We then use a separation of time scales between fast co-ordinates ($\alpha_k$) and the slow one ($\beta$).
 
Recall that for low temperatures we have $\alpha_k= O(T^{1/2})$ and we assume $\varepsilon = O(T)$.
We expand all terms in the Fokker-Planck Eq.~\eqref{suppeq:fp-curvi} up to linear order in $T$.  For the metric itself, we obtain
\begin{align}
\begin{split}
&T g^{nm} =  T \left(
\begin{array}{cccccccc}
 1 & 0 & 0 & 0 & 0 & 0 & 0 & 0 \\
 0 & 1 & 0 & 0 & 0 & 0 & 0 & 0 \\
 0 & 0 & \frac{1}{2} & 0 & 0 & 0 & 0 & 0 \\
 0 & 0 & 0 & 2  & 0 & 0 & 0 & 0 \\
 0 & 0 & 0 & 0 & 1 & 0 & 0 & 0 \\
 0 & 0 & 0 & 0 & 0 & 1 & 0 & 0 \\
 0 & 0 & 0 & 0 & 0 & 0 & 1 & 0 \\
 0 & 0 & 0 & 0 & 0 & 0 & 0 & 1 \\
\end{array}
\right)  + O(T^{3/2})  
\end{split}
\end{align}
corresponding to Gaussian white noise acting independently on each coordinate.
The generalised force is instead given by
\begin{equation}
    \mu_m(\bm{q})  
    = \sum_i F_i  (\bm{X}(\bm{q}))  J_{im} .
    \label{drift fokker planck}
\end{equation}
where $F_i$ is a Cartesian component of the force.  
To make use of the formulae \eqref{eq:eff_V_spring} and \eqref{eq:eff_V_angle} for the potential, it is useful to express the force in terms of gradients with respect to the curvilinear co-ordinates.  That is,
\begin{equation}
    F_i = - \sum_j (\delta_{ij} + \frac{k_a}{k} \varepsilon_{ij}) \frac{\partial V}{ \partial X_j}  = - \sum_{j,n}(\delta_{ij} + \frac{k_a}{k} \varepsilon_{ij}) \frac{\partial q_n }{ \partial X_j} \frac{\partial V} { \partial q_n}  = -  \sum_{j,n} (\delta_{ij} + \frac{k_a}{k} \varepsilon_{ij})\frac{\partial V}{ \partial q_n}  (J^{-1})_{nj},
\end{equation}
where $\varepsilon_{ij}$ is 2D Levi-Civita symbol and $V = V_{\rm spring} + V_{\rm angle}$, as given by the  \eqref{eq:eff_V_spring} and \eqref{eq:eff_V_angle}. 
Hence
\begin{equation}
    \mu_m = -\frac{\partial V}{ \partial q_m} 
    + \frac{k_a}{k}  \sum_{i,j,n} \frac{\partial V}{ \partial q_n}  (J^{-1})_{nj} \varepsilon_{ji}   J_{im} 
\end{equation}
As expected, these (odd) forces are not a pure gradient unless  $k_a = 0$. 

The Jacobian of the nonlinear coordinate transformation is given explicitly (for $\theta=0$) by
\begin{equation}
\Matrix{J} = 
\begin{pmatrix}
\frac{1}{2} & 0 & \frac{\alpha_e}{2} + \frac{\alpha_c \sin\!\left(\frac{\beta}{2}\right)}{\sqrt{2}} & \frac{(-\sqrt{2}+\alpha_d)\sin\!\left(\frac{\beta}{2}\right)}{2\sqrt{2}} & 0 & -\frac{\cos\!\left(\frac{\beta}{2}\right)}{\sqrt{2}} & 0 & -\frac{1}{2} \\[6pt]
0 & \frac{1}{2} & -\frac{\alpha_f}{2} - \frac{(-\sqrt{2}+\alpha_d)\cos\!\left(\frac{\beta}{2}\right)}{\sqrt{2}} & -\frac{\alpha_c \cos\!\left(\frac{\beta}{2}\right)}{2\sqrt{2}} & -\frac{\sin\!\left(\frac{\beta}{2}\right)}{\sqrt{2}} & 0 & -\frac{1}{2} & 0 \\[6pt]
\frac{1}{2} & 0 & -\frac{\alpha_e}{2} + \frac{(-\sqrt{2}+\alpha_d)\sin\!\left(\frac{\beta}{2}\right)}{\sqrt{2}} & \frac{\alpha_c \sin\!\left(\frac{\beta}{2}\right)}{2\sqrt{2}} & -\frac{\cos\!\left(\frac{\beta}{2}\right)}{\sqrt{2}} & 0 & 0 & \frac{1}{2} \\[6pt]
0 & \frac{1}{2} & \frac{\alpha_f}{2} - \frac{\alpha_c \cos\!\left(\frac{\beta}{2}\right)}{\sqrt{2}} & -\frac{(-\sqrt{2}+\alpha_d)\cos\!\left(\frac{\beta}{2}\right)}{2\sqrt{2}} & 0 & -\frac{\sin\!\left(\frac{\beta}{2}\right)}{\sqrt{2}} & \frac{1}{2} & 0 \\[6pt]
\frac{1}{2} & 0 & \frac{\alpha_e}{2} - \frac{\alpha_c \sin\!\left(\frac{\beta}{2}\right)}{\sqrt{2}} & -\frac{(-\sqrt{2}+\alpha_d)\sin\!\left(\frac{\beta}{2}\right)}{2\sqrt{2}} & 0 & \frac{\cos\!\left(\frac{\beta}{2}\right)}{\sqrt{2}} & 0 & -\frac{1}{2} \\[6pt]
0 & \frac{1}{2} & -\frac{\alpha_f}{2} + \frac{(-\sqrt{2}+\alpha_d)\cos\!\left(\frac{\beta}{2}\right)}{\sqrt{2}} & \frac{\alpha_c \cos\!\left(\frac{\beta}{2}\right)}{2\sqrt{2}} & \frac{\sin\!\left(\frac{\beta}{2}\right)}{\sqrt{2}} & 0 & -\frac{1}{2} & 0 \\[6pt]
\frac{1}{2} & 0 & -\frac{\alpha_e}{2} - \frac{(-\sqrt{2}+\alpha_d)\sin\!\left(\frac{\beta}{2}\right)}{\sqrt{2}} & -\frac{\alpha_c \sin\!\left(\frac{\beta}{2}\right)}{2\sqrt{2}} & \frac{\cos\!\left(\frac{\beta}{2}\right)}{\sqrt{2}} & 0 & 0 & \frac{1}{2} \\[6pt]
0 & \frac{1}{2} & \frac{\alpha_f}{2} + \frac{\alpha_c \cos\!\left(\frac{\beta}{2}\right)}{\sqrt{2}} & \frac{(-\sqrt{2}+\alpha_d)\cos\!\left(\frac{\beta}{2}\right)}{2\sqrt{2}} & 0 & \frac{\sin\!\left(\frac{\beta}{2}\right)}{\sqrt{2}} & \frac{1}{2} & 0
\end{pmatrix}
\end{equation}
It is sufficient to restrict the following analysis to the case $\theta=0$ because rotational invariance means that the dynamics of the $\alpha$ and $\beta$ co-ordinates are independent of $\theta$.


All together, the three drift terms are given, to leading order in $T$, by
\begin{align}
\begin{split}
    g^{nm} \mu_m  \approx\ 
    k\begin{pmatrix}
        0 \\
        0 \\
        -2 \alpha_e \alpha_f  \cos \beta \\
        2 (\alpha_f^2 - \alpha_e^2) \sin \beta \\
        -2 \alpha_c  \\
        -2 \alpha_d  \\
        2 \alpha_e  (\cos \beta - 1) \\
        -2 \alpha_f  (\cos \beta + 1)
\end{pmatrix}
    + 
    k_a\begin{pmatrix}
        0 \\
        0 \\
         -\left(\alpha_c^2 + \alpha_d (\alpha_d + \sqrt{2})\right) \\
        -2 \alpha_c (2 \alpha_d + \sqrt{2})  \\
        \sqrt{2}  (-\alpha_f^2 + \alpha_e^2) \sin \beta \\
        2 \sqrt{2} \alpha_e \alpha_f  \cos \beta \\
        \alpha_f  (\sqrt{2} \alpha_d + 2 \cos \beta + 2) \\
        \alpha_e  (-\sqrt{2} \alpha_d + 2 \cos \beta - 2)
\end{pmatrix}
+
\frac{2\varepsilon}{k}
\left(
\begin{array}{c}
 0 \\
 0 \\
 0 \\
 - 2k\sin (4 \beta)\\
   k_a \sqrt{2}\sin (4 \beta) \\
 0 \\
 0 \\
 0\\
\end{array}
\right)
+ O(T^{3/2})
\end{split}
\end{align}
also,
\begin{align}
T \frac{\partial }{\partial q_m} g^{nm} = O(T^{3/2}) 
\end{align}
and
\begin{align}
\begin{split}
\frac{1}{2}T   g^{nm} \frac{\partial }{\partial q_m} \log (\det g) =T   g^{nm} \frac{\partial }{\partial q_m} \log (\det|\Matrix{J}|) \approx 
    \begin{pmatrix}
        0 \\
        0 \\
        0 \\
        0 \\
        0 \\ 
        -T\sqrt{2} \\ 
        0 \\
        0
    \end{pmatrix} + O(T^{3/2})
\end{split}
\end{align}
Summing all contributions and switching to the Langevin representation, the low-temperature dynamics of the square model in curvilinear coordinates is:
\begin{multline}
\begin{pmatrix}
        \rlj{2}\dot{x}_{\rm CM} \\
        \rlj{2}\dot{y}_{\rm CM} \\
        \dot{\theta} \\
        \dot{\beta} \\
        \dot{\alpha_c} \\
        \dot{\alpha_d}  \\
        \dot{\alpha_e} \\
        \dot{\alpha_f}
\end{pmatrix}
=
k\begin{pmatrix}
        0 \\
        0 \\
        -2 \alpha_e \alpha_f \cos\beta \\
        2 (\alpha_f^2 - \alpha_e^2)\sin\beta \\
        -2 \alpha_c \\
        -2 \alpha_d \\
        2 \alpha_e(\cos\beta - 1) \\
        -2 \alpha_f(\cos\beta + 1)
\end{pmatrix}
+
k_a\begin{pmatrix}
        0 \\
        0 \\
        -\alpha_c^2 - \alpha_d(\alpha_d+\sqrt{2}) \\
        -2\alpha_c(2\alpha_d+\sqrt{2}) \\
        \sqrt{2}(-\alpha_f^2+\alpha_e^2)\sin\beta \\
        2\sqrt{2}\alpha_e\alpha_f\cos\beta \\
        \alpha_f(\sqrt{2}\alpha_d+2\cos\beta+2) \\
        \alpha_e(-\sqrt{2}\alpha_d-2+2\cos\beta)
\end{pmatrix}
+
T\begin{pmatrix}
        0 \\
        0 \\
        0 \\
        0 \\
        0 \\
        -\sqrt{2} \\
        0 \\
        0
\end{pmatrix}
\\
+
\frac{2\varepsilon}{k}
\left(
\begin{array}{c}
 0 \\
 0 \\
 0 \\
 -2k \sin (4 \beta)\\
   k_a\sqrt{2} \sin (4 \beta) \\
 0 \\
 0 \\
 0\\
\end{array}
\right)
+\sqrt{2T} \left(
\begin{array}{c}
\eta^{x_{\rm CM}}\\
\eta^{y_{\rm CM}}\\
\eta^{\theta}/ \sqrt{2}   \\
 \eta^{\beta} \sqrt{2} \\  
 \eta^{\alpha_c}\\
 \eta^{\alpha_d}\\
 \eta^{\alpha_e}\\
 \eta^{\alpha_f}\\
\end{array}
\right).
\label{full dynamics}
\end{multline}

\subsubsection{Adiabatic elimination and effective folding dynamics}

As already noted, the deformation co-ordinates $\alpha$ are stiff, they values are $O(T^{1/2})$ at low temperatures.  By considering the passive dynamics, one also sees that they relax quickly, on a time scale of order $k^{-1} =O(1)$.  On the other hand, the  $\beta$ co-ordinate is soft and has slow dynamics in the passive system, with a time scale of order $\varepsilon^{-1}$ from the angular potential, and of order $T^{-1}$ from diffusion (recall we assumed that $\varepsilon \sim T$ so these time scales are of the same order). This separation of time scales is preserved in the presence of transverse forces, $k_a\neq 0$.  As explained in the main text, the elimination of the fast variables can be achieved by implementing the general method of \cite{pavliotis2008multiscale}, see Ch. 11. 
We outline here a physical derivation.

It is important that $\alpha_c$ appears linearly in the equation of motion for $\beta$, but the other co-ordinates $\alpha_{d,e,f}$ all appear only in second-order terms.  
This allows for an adiabatic factorisation of the joint probability distribution,
\begin{equation}
    P(\beta, \theta, \bm{\alpha};t) \approx P_s(\beta, \theta, \alpha_c;t) P_q(\alpha_d, \alpha_e, \alpha_f| \beta, \theta, \alpha_c),
\end{equation}
where we recall that $\theta$ indicates the global rotation of the system. For the fast modes, the conditional distribution is equilibrium-like and given by
\begin{equation}
    P_q(\alpha_d, \alpha_e, \alpha_f| \beta, \theta, \alpha_c)  = Z_q ( \beta, \theta, \alpha_c; T)^{-1} |J|e^{-\frac{V(\beta, \alpha_d, \alpha_e, \alpha_f)}{T}},
    \label{quick modes distribution}
\end{equation}
with $Z_q$ the normalisation factor and $|J|$ the Jacobian determinant as given in Eq.~\eqref{eq:jacob_square}. 

To obtain the marginal dynamics for the internal angle $\beta$ only, we integrate \eqref{full dynamics} with respect to the conditional density $P_q(\alpha_d,\alpha_e,\alpha_f|\beta,\theta,\alpha_c)$ of the fast modes, as given in Eq.~\eqref{quick modes distribution}, effectively replacing every term that depends on the fast variables by its conditional average, which we denote $\langle \cdot \rangle_\beta$ for compactness.
In particular, $\langle \alpha_n \alpha_m \rangle_{\beta} = 0 $ for $n \neq m$ and $n,m \in \{d,e,f\}$, while $\langle \alpha_n \rangle_{\beta} =0$ for $n \in \{e,f\}$ and $\langle \alpha_d \rangle_{\beta} =O(T)$.
In fact, the only non-trivial average is:
\begin{equation}
    \left\langle \alpha_f^2-\alpha_e^2\right\rangle_{\beta} = -\frac{T \cos \beta}{k \sin^2\beta }.
\end{equation}
The adiabatically-averaged coupled Langevin dynamics for $\beta$ and $\alpha_c$ are then given by
\begin{align}
\begin{split}
\begin{pmatrix}
        \dot{\beta} \\
        \dot{\alpha_c} \\
\end{pmatrix}
=
&
\begin{pmatrix}
        -2 T \cot \beta \\
        -2k \alpha_c  \\
\end{pmatrix}
    + 
\begin{pmatrix}
        -2 \sqrt{2} \alpha_c k_a  \\
        \frac{\sqrt{2} T k_a \cot \beta}{k} \\
\end{pmatrix}
+
\varepsilon
\begin{pmatrix}
        -4 \sin (4 \beta )  \\
        0 \\
\end{pmatrix}
+ \frac{\varepsilon k_a}{k}
\begin{pmatrix}
        0  \\
        2 \sqrt{2} \sin (4 \beta ) \\
\end{pmatrix}
+
\begin{pmatrix}
        2\sqrt{T } \eta^{\beta} \\
        \sqrt{2T} \eta^{\alpha_c} \\
\end{pmatrix}.
\label{supp Beta and alphac dynamics approx}
\end{split}
\end{align}
As already discussed in Methods \textbf{A.4}, a further coarse-graining step finally gives the reduced equation 
\begin{equation}
    \dot\beta = -2T \Lambda\cot{\beta}-4 \varepsilon \Lambda \sin (4 \beta) + 2\sqrt{T \Lambda} \eta,
\end{equation}
where $\eta$ is a unit Gaussian white noise and $\Lambda = 1 + {k_a^2}/{k^2}$ is the activity-induced speedup factor .

\subsection*{Numerical methods } \label{sm:numerics}

\subsubsection*{Timestep choice}
To ensure numerical stability, the integration time step was set to:
\begin{equation}
\Delta t = dt_{\mathrm{coef}} \min\left\{\frac{1}{k},\frac{1}{k_a},\,\frac{1}{2\xi_{\mathrm{max}}^2 T}\right\},
\end{equation}
where $dt_{\mathrm{coef}}<1$ is a dimensionless safety factor. In the absence of activity ($k_a=0$), the minimum was evaluated over the remaining terms only.

Unless stated otherwise, simulations in \textbf{Fig.1} and \textbf{Fig.2} and \textbf{Supplementary video 1} used $dt_{\mathrm{coef}}=0.05$. Two exceptions were considered: in \textbf{Fig.1e} (top panel), $dt_{\mathrm{coef}}=0.0005$ was employed due to the short assembly timescale ($\sim10^{-1}$), and in \textbf{Fig.2a}, $dt_{\mathrm{coef}}=0.1$ was sufficient given the finite resolution of the heatmap.

For \textbf{Fig.3} and \textbf{Supplementary video 2}, a smaller value $dt_{\mathrm{coef}}=0.025$ was adopted to accurately resolve the steep repulsive regime of the Morse potential at short interparticle distances.

Ensemble averages were computed over $N$ statistically independent trajectories. Error bars represent the standard error of the mean, obtained as the sample standard deviation divided by $\sqrt{N}$. When multiple events were extracted from a single trajectory, measurements were initiated only after an initial transient period, allowing the system to reach its steady-state regime.

\subsubsection*{Force regularization}

To ensure numerical stability, short-distance regularisations were applied to both pairwise and angular interactions.

For the pairwise spring forces, unit vectors were computed as $\hat{\mathbf r}_{ij} = \mathbf r_{ij}/r_{ij}$, with the convention $\hat{\mathbf r}_{ij}=0$ for $r_{ij}=0$. This avoids singularities in the force evaluation when two particles transiently overlap due to noise. This convention was used in all simulations (all figures and Supplementary Movies).

For the three-body angular interaction (used only in \textbf{Fig.2c}), the force is derived from the angle-dependent potential:
\begin{equation}
    V_{\rm angle} (\bm{X})= -\frac{\varepsilon}{4}\sum_{i=1}^4  \left[ 2\left (\frac{\bm{r}_{i,i-1}\cdot\bm{r}_{i+1,i}}{|\bm{r}_{i,i-1}||\bm{r}_{i+1,i}|}  \right )^2-1\right]^2,
\end{equation}
which depends explicitly on distances between the particles $\bm{r}_{i,k}$. As a result, the corresponding forces contain the inverse of the particle distances and become singular when adjacent particles approach vanishing separation. To regularise this, the angular contribution was set to zero whenever either of the bond lengths in a triplet fell below a small cutoff. This prevents numerical instabilities associated with ill-defined bond angles at short distances.

\subsubsection*{Numerical computations of the Stratonovich integral}
As mentioned in the Methods \textbf{A.5}, work is defined via a Stratonovich integration rule. Numerically, the Stratonovich integral is defined by a midpoint rule:
\begin{equation}
W \approx \sum_{n,i} 
\bm{F}_i \left(\tfrac{1}{2}\left[\bm{r}_i(t_{n+1})+\bm{r}_i(t_n)\right]\right)
\cdot \left[\bm{r}_i(t_{n+1})-\bm{r}_i(t_n)\right],
\end{equation}
where the sum runs over discrete time steps $t_n=n\Delta t$ and over all particles. The force entering the midpoint evaluation includes both reciprocal and non-reciprocal contributions.

\subsubsection*{Self-assembly with cargo}
The double-square system follows the dynamics described in Methods \textbf{A.2}.  To simulate the system-wide stabilization shown in Fig.~\textbf{3}\textbf{d}–\textbf{e} of the main text, we used the following protocol:
\begin{enumerate}
    \item Initialize the assembly in the rectangular configuration (a special case of the ladder state) and measure the time required to reach the chevron configuration, as defined in Methods~\textbf{A.6}.
    \item Repeat for $N$ statistically independent realizations.
    \item Collect and sort the transition times to construct the density plot in Fig.~\textbf{3}\textbf{e}, representing the distribution of first-passage times from ladder to chevron.
\end{enumerate}
This protocol assumes that once the chevron configuration is reached, the double-square is immediately stabilized by a bound cargo and does not revert to the ladder state. The density plot therefore quantifies the distribution of transition times under thermal fluctuations and transverse forces.


\end{document}